\documentclass[sigconf]{acmart}
\usepackage{graphicx}
\usepackage{array}
\usepackage{subfig}
\usepackage{multirow}
\usepackage{booktabs}

\AtBeginDocument{%
  \providecommand\BibTeX{{%
    \normalfont B\kern-0.5em{\scshape i\kern-0.25em b}\kern-0.8em\TeX}}}

\copyrightyear{2025}
\acmYear{2025}
\setcopyright{cc}
\setcctype{by-nc-nd}
\acmConference[CHI '25]{CHI Conference on Human Factors in Computing Systems}{April 26-May 1, 2025}{Yokohama, Japan}
\acmBooktitle{CHI Conference on Human Factors in Computing Systems (CHI '25), April 26-May 1, 2025, Yokohama, Japan}\acmDOI{10.1145/3706598.25713265}
\acmISBN{979-8-4007-1394-1/25/04}

\begin{document}

\title[Exploring DIY Smart Home Building Experiences with VLM-Based Camera Sensors]{“What If Smart Homes Could See Our Homes?”: Exploring DIY Smart Home Building Experiences with VLM-Based Camera Sensors}

\author{Sojeong Yun}
\affiliation{%
  \institution{Department of Industrial Design, KAIST}
  \city{Daejeon}
  \country{Republic of Korea}}
\email{thwjd3785@kaist.ac.kr}

\author{Youn-kyung Lim}
\affiliation{%
  \institution{Department of Industrial Design, KAIST}
  \city{Daejeon}
  \country{Republic of Korea}}
\email{younlim@kaist.ac.kr}

\renewcommand{\shortauthors}{Yun,S. and Lim,Y.}

\begin{abstract}
    The advancement of Vision-Language Model (VLM) camera sensors, which enable autonomous understanding of household situations without user intervention, has the potential to completely transform the DIY smart home building experience. Will this simplify or complicate the DIY smart home process? Additionally, what features do users want to create using these sensors? To explore this, we conducted a three-week diary-based experience prototyping study with 12 participants. Participants recorded their daily activities, used GPT to analyze the images, and manually customized and tested smart home features based on the analysis. The study revealed three key findings: (1) participants’ expectations for VLM camera-based smart homes, (2) the impact of VLM camera sensor characteristics on the DIY process, and (3) users’ concerns. Through the findings of this study, we propose design implications to support the DIY smart home building process with VLM camera sensors, and discuss living with intelligence.
\end{abstract}



\maketitle

\section{Introduction}

\textcolor{black}{Do-It-Yourself (DIY)} smart home systems enable users to freely combine sensors and actuators to directly build smart home features that meet their specific needs \cite{woo2015user}. Since the meaning of a home and the values emphasized within that space differ from person to person, the needs for smart home features also vary across users \cite{tao2021diy, klapperich2020designing, chiang2020exploring, williams2020upcycled}. In this context, DIY smart homes fully respect users' autonomy by providing the flexibility to build the features they require \cite{woo2015user, williams2019understanding, aiswarya2021living, verweij2019exploring, jakobi2018evolving, de2019user}. To date, numerous studies on DIY smart homes have explored various approaches to support users' DIY practices \cite{reisinger2017visual}, but despite these efforts, the complexity of the construction process remains a major barrier to the widespread adoption of DIY smart homes \cite{he2019smart, jakobi2017catch}. To address these limitations, prior research has focused on developing technologies such as natural language-based processes for building smart home features \cite{liu2023understanding, king2024sasha} and tools that support seamless debugging \cite{zaidi2023user, zhang2023helping, coppers2020fortniot}, demonstrating efforts to shift the DIY smart home experience from a technology-centered approach to a user-centered one. 

One of the significant turning points in transforming the DIY smart home building experience into a user-centered process is the advancement of Vision-Language Model (VLM) technology, which enables machines to visually perceive the world in a human-like manner \cite{gonzalez2024investigating, team2023gemini, yun2024toolkit}. Human vision is known to account for over 80\% of the sensory information used to perceive the external world \cite{ripley2010vision}, suggesting that machines equipped with cameras could potentially perceive the world similarly to humans. Previous studies have explored ways to use home cameras as sensors, but conventional ML-intensive camera sensors could not automatically recognize the contextual situations of a home environment. Users were required to manually synthesize information from objects, human poses, facial expressions, and gestures within the scene to interpret and define the meaning of situations \cite{yun2023potential}. In contrast, VLM-based camera sensors can deeply understand home contexts in a human-like way, eliminating the need for users to define situations themselves and significantly reducing the manual sense-making effort \cite{moore2018managing, kurze2020guess}. These advancements open up new possibilities for users to customize smart home features more easily and flexibly. \textcolor{black}{For example, the VLM camera sensor can autonomously detect whether a person is cooking, washing dishes, or taking vitamins in the kitchen, so that users can easily set up automation features tailored to each situation.} However, despite the unique potential of VLM camera sensor-based DIY smart homes, no studies have yet explored users' expectations for these systems or examined how the unique characteristics of VLM camera sensors might introduce new possibilities and challenges for DIY smart home development. 

\textcolor{black}{Building on this background, our research goal is to explore what users expect from a VLM camera sensor-based smart home and how their experiences differ from traditional DIY processes when implementing their expectations in a DIY manner.} As an early-stage exploration of this potential, we aimed to uncover diverse user expectations and experiences unconstrained by technical limitations. To achieve this, we proposed a diary-based DIY toolkit concept aimed at helping users maintain their existing mental models of DIY smart home construction and conducted a three-week experience prototyping study using this diary. Twelve participants filmed their daily activities, observed how the footage was analyzed, and engaged in a simulated DIY process of defining and testing smart home features. From this exploration, we identified: (1) participants’ expectations for VLM camera-based smart homes (Section 4.1), (2) the impact of VLM camera sensor characteristics on the DIY process (Section 4.2), and (3) concerns including privacy issues (Section 4.3). Based on these findings, we propose design implications to support the VLM camera sensor-based DIY smart home building process (Section 5.1) and discuss living with intelligence (Section 5.2, 5.3). \textcolor{black}{Our key contribution in this study is that we first proposed and identified user-centered VLM camera sensor-based DIY smart home systems.}

\section{Related Work}
\subsection{Emerging Smart Services with VLM}

The advancement of AI technologies has advanced VLMs' ability to interpret and describe images based on user requests. These improvements in image analysis enable the contextual and detailed description of visual scenes, allowing for detailed responses to visual information-based inquiries. This technological progress has facilitated the development of various camera-centered smart services, expanding possibilities in areas such as smart homes.

Previous studies have explored the potential of converting visual information into audio, developing tools to assist blind or low-vision individuals with cooking \cite{li2024contextual}, watching mobile content \cite{ning2024spica, liu2024unblind}, and locating personal belongings \cite{wen2024find}. Additionally, research has examined user needs for AI-powered scene description applications to support blind individuals \cite{gonzalez2024investigating}. Some studies have gone beyond scene description, creating systems that answer questions based on medical images \cite{yildirim2024multimodal} or developing mobile AR systems to inspect indoor safety issues and offer advice for mitigation \cite{su2024rassar}. Notably, according to Google's Gemini document \cite{team2023gemini}, features such as plant care guidance, recipe suggestions based on available ingredients, and soccer technique improvement through image analysis are feasible. These advancements in image analysis technology enable smart services to visually interpret home situations and assist users autonomously.

Expanding this potential to the smart home domain, \textcolor{black}{VLM camera sensors allow users to build various smart home features that were previously unattainable, offering new possibilities for DIY smart home construction. However, despite this potential, no research has yet explored these experiences. In response to these research gaps, our study examines what roles users want smart home features to play when built with VLM camera sensors and how we can support users' DIY process.}

\subsection{DIY Smart Home Building Process}

The value of DIY smart homes lies in the autonomy it provides users to create features tailored to their specific needs \cite{yun2023potential, abdallah2017lessons, woo2015user, verweij2019exploring}. Users can design their own smart home services using sensors and actuators in a DIY manner. This process typically follows the Trigger-Action paradigm \cite{brackenbury2019users, ur2014practical, zhao2021understanding}, involving two main steps: \textbf{1) defining the conditions for actions} by gathering sensor data, and \textbf{2) establishing rules for actuators to perform actions} when those conditions are met. Users first collect data from home sensors, then engage in a sense-making process to identify key sensor values that represent conditions for automation \cite{moore2018managing, kurze2020guess}. Afterward, users create automation rules using an “If-this-then-that” structure \cite{huang2015supporting, yu2021analysis}. Once the rules are defined, it is essential to test and refine the feature to ensure it functions according to the user’s specifications.

While VLM camera sensors based on visual sensing cannot fully replace commonly used IoT sensors responsible for other sensory data such as temperature and humidity, their ability to contextually understand household situations through visual elements alone opens up significant possibilities, particularly in providing a vast range of triggers that can be used for automation \cite{yun2023potential, yeo2023omnisense, li2019fmt}. Particularly, since VLM sensors can autonomously understand household situations \cite{team2023gemini}, they can bypass the sense-making process—interpreting what the sensor values mean—a key process of traditional DIY smart home setups. \textcolor{black}{While this capability may both simplify the process of building a DIY smart home and introduce new challenges, these aspects have not yet been studied. In response to this research gap,} this study examines the advancements of these sensors, focusing on user experiences in creating smart home features with VLM camera sensors in a DIY context, and explores both the opportunities presented and the additional support that may be required.

\section{Method}

To determine the most suitable method for observing the DIY smart home building experience with the introduction of VLM camera sensors, we reviewed prior research on smart homes. Previous studies exploring user experiences with smart homes have employed the following three approaches to understanding how users interact with smart homes: 1) implementing and deploying a fully functional system \cite{woo2015user, kim2022exploration}, 2) creating a controlled lab environment and conducting a wizard-of-oz study with researcher involvement \cite{liu2023understanding}, and 3) designing simulated experiences without deploying an actual system \cite{yun2023potential}.

In selecting the appropriate study method, our primary concerns were: a) protecting participants' sensitive data, and b) ensuring participants could fully experience the process of building and testing their DIY smart home ideas. Based on these considerations, we ruled out the first method, as deploying a functional system could risk indiscriminate exposure of sensitive image data within participants’ homes. Similarly, we excluded the second method, as a lab environment may not capture the unique and unpredictable situations that occur in real homes, limiting the ability to fully observe the DIY building process.

The final method we considered involved designing a study where participants could simulate the DIY smart home-building process without deploying a real system. This approach appeared the most practical, as it enabled participants to experience the DIY process within their own homes while minimizing data exposure. However, given the study's objectives, it was crucial to ensure participants could engage with every detailed step of the DIY smart home construction, even in a simulated context. Ultimately, we selected a diary-based experience prototyping method (Section 3.1). The following section will detail the specific considerations taken into account when designing this study.

\subsection{Experience Prototyping Study Design}

Our goal was to design a study that allowed participants to simulate all the key experiential elements they would encounter during an actual DIY smart home-building process. To achieve this, we adopted a diary-based experience prototyping approach \cite{buchenau2000experience}, guiding participants through the entire experience of constructing DIY smart home features.

The essential experiential elements in the user’s DIY smart home building process can be divided into two key stages \cite{ur2014practical, zhao2021understanding}. The first stage involves placing sensors and receiving sensor data, while the second stage is about defining rules and testing them. For the first stage, placing camera sensors and receiving data, we aimed to provide a realistic experience even if real-time analysis of the home environment was not feasible. To simulate this, participants were asked to record videos within their homes, select key scenes from the footage, and process important information by blurring sensitive details. Then, using ChatGPT, participants would analyze the image. One challenge in this step was accounting for the wide variety of home camera forms, which meant that we needed to provide participants with equipment that could simulate this diversity. For the second stage, planning and testing features, we aimed to leverage the possibility of having participants input functional rules and images as prompts into ChatGPT, which could then generate commands capable of operating a real smart home system. Since completing the prompts in line with the actual DIY smart home building process could be complex, we recognized the need to offer participants step-by-step guidance throughout the prompt creation process. Based on these conclusions, we determined that two key materials were necessary for this study.

\textbf{1) Categorizing Various Home Camera Configurations:} First, we aimed to categorize how camera sensors could be applied in various forms within homes. From reviewing previous studies that utilized home cameras as sensors, we found that wearable cameras effectively captured users' interactions with objects \cite{li2019fmt, li2024contextual, lee2024gazepointar}, cameras embedded in home robots could efficiently recognize full-body poses of individuals \cite{wang2024pepperpose}, mobile-mounted cameras allowed for flexible indoor sensing while freely moving through indoor spaces \cite{wen2024find, su2024rassar}, and home security cameras were able to continuously monitor the residents \cite{cho2023ai}. Given that different types of home cameras provide distinctive sensing capabilities, we considered that the form of camera sensors could influence participants' ideation for smart home features. Therefore, by considering this diversity in camera sensors, we sought to compile camera sensor configurations that could offer meaningful differences in sensing roles. This process will be further detailed in Section 3.1.1.

\textbf{2) Designing a Toolkit to Support DIY Process:} In designing a diary-style toolkit, it was crucial to develop guides that would help participants effectively compose the final prompt, taking into account all the elements they considered during their existing DIY smart home building process. To design these guides, we referred to prior studies that summarized the DIY setup process \cite{yun2023potential, woo2015user, huang2015supporting, yu2021analysis}. This process will be elaborated further in Section 3.1.2.

\subsubsection{Proposing Three Camera Sensor Configurations}

We aimed to categorize existing home cameras under the assumption that cameras with similar functionalities would perform similar sensing roles. A prior study proposed three classification criteria: (1) Does the camera stay or move? (2) Where does it sense? (3) How often does it sense? \cite{pierce2022addressing}. While that study categorized cameras by whether they sense inside or outside, we focused on indoor usage and refined the second criterion to, “Does the camera focus on a single area or multiple areas simultaneously?” Additionally, we decided that criterion 3 (“How often does it sense?”) is influenced more by automation rules than a camera type. Therefore, this will be addressed in Section 3.1.2 where we discuss how participants created sensing rules to sense home environments only when necessary. Using the first two criteria, we classified cameras into four types: (1) Moving cameras with limited coverage, (2) Moving cameras with wide coverage, (3) Fixed cameras with wide coverage, and (4) Fixed cameras with limited coverage. Finally, we mapped various camera forms mentioned in previous studies on home cameras \cite{pierce2022addressing, li2019fmt, li2024contextual, lee2024gazepointar, wang2024pepperpose, wen2024find, su2024rassar, cho2023ai, pierce2019smart, hu2025elegnt} to these four types as follows. 

Type (1), moving cameras with limited coverage, includes wearable cameras or those attached to home robots designed to track specific subjects. Type (2), moving cameras with wide coverage, consists of 360-degree rotating cameras capable of monitoring multiple areas. Type (3), fixed cameras with wide coverage, achieves broader coverage by employing multiple cameras that collectively monitor different spaces. Lastly, type (4), fixed cameras with limited coverage, focuses on specific devices or spaces, such as a refrigerator or table lamp. We considered that types (2) and (3) serve a similar purpose—comprehensive monitoring of wide spaces—either through a single rotating camera or multiple fixed cameras. Therefore, we merged them into a single category, consolidating camera sensor types into three configurations. \textcolor{black}{Figure 1 illustrates the process leading to the derivation of three camera sensor configuration concepts.}

\begin{figure}
    \centering
    \includegraphics[width=1\linewidth]{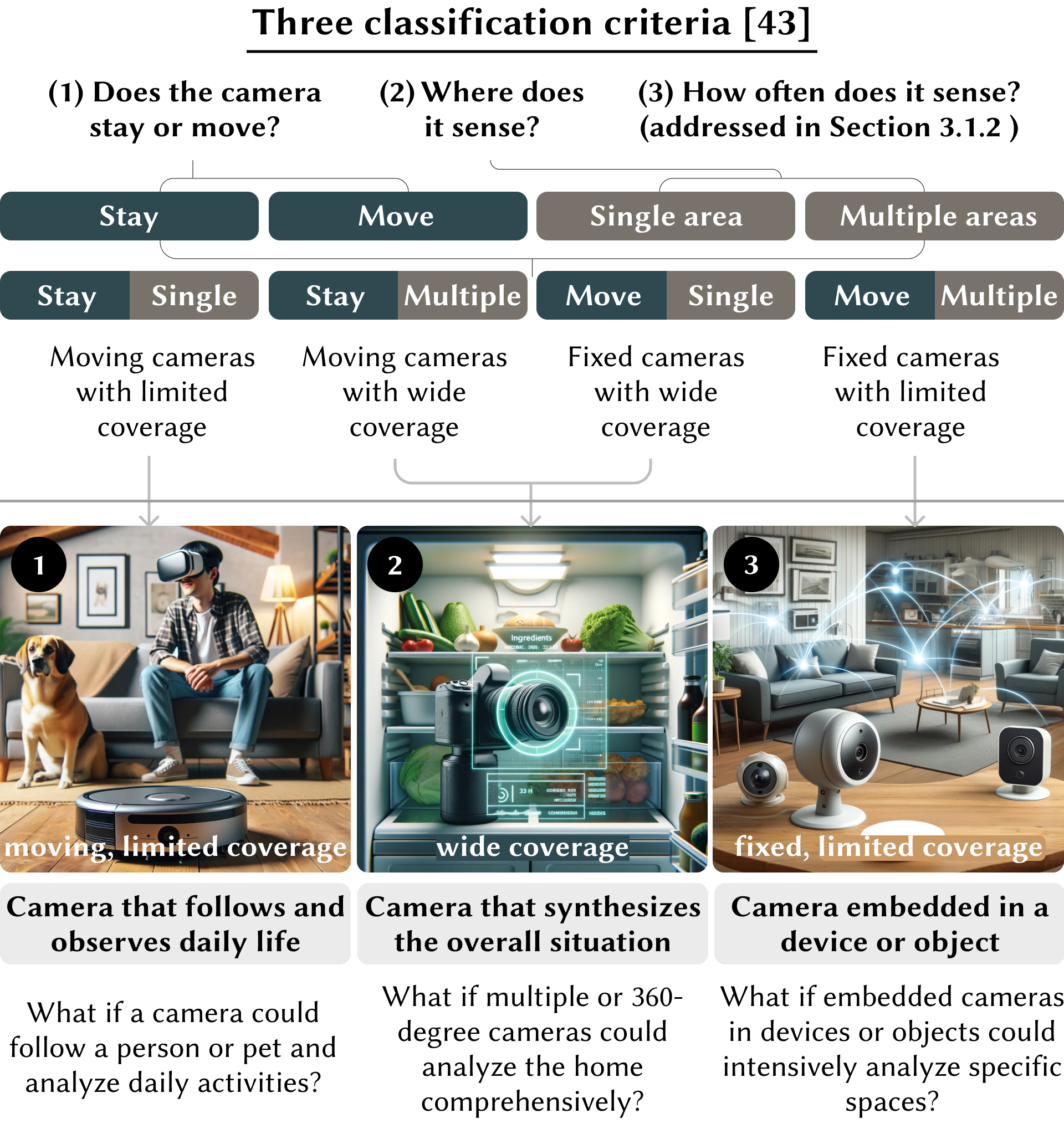}
    \caption{Three proposed concepts  \textcolor{black}{and the process of deriving them} (All images were created using DALL-E 3)}
    \Description{(a) Concept 1. Camera that follows and observes daily life: What if a camera could follow a person or pet and analyze daily activities?, (b) Concept 2. Camera that synthesizes the overall situation: What if multiple or 360-degree cameras could analyze the home comprehensively?, (c) Concept 3. Camera embedded in a device or object: What if embedded cameras in devices or objects could intensively analyze specific spaces?}
    \label{fig:enter-label}
\end{figure}

In this study, we aimed to encourage participants to generate various DIY smart home feature ideas based on these three configurations. We provided GoPros with various mounting accessories, allowing participants to experience all three camera configurations and easily position the cameras wherever they preferred. Since GoPros do not have an automatic rotation function, we also provided a 360-degree rotating turntable for use. Additionally, if participants required multiple cameras, we encouraged them to use their mobile devices. For scenarios requiring a moving camera, we suggested that they ask family members or friends to help with filming. Figure 2 illustrates examples of participants using cameras, mounts, and turntables to implement various camera configurations.

\begin{figure}
    \centering
    \includegraphics[width=1\linewidth]{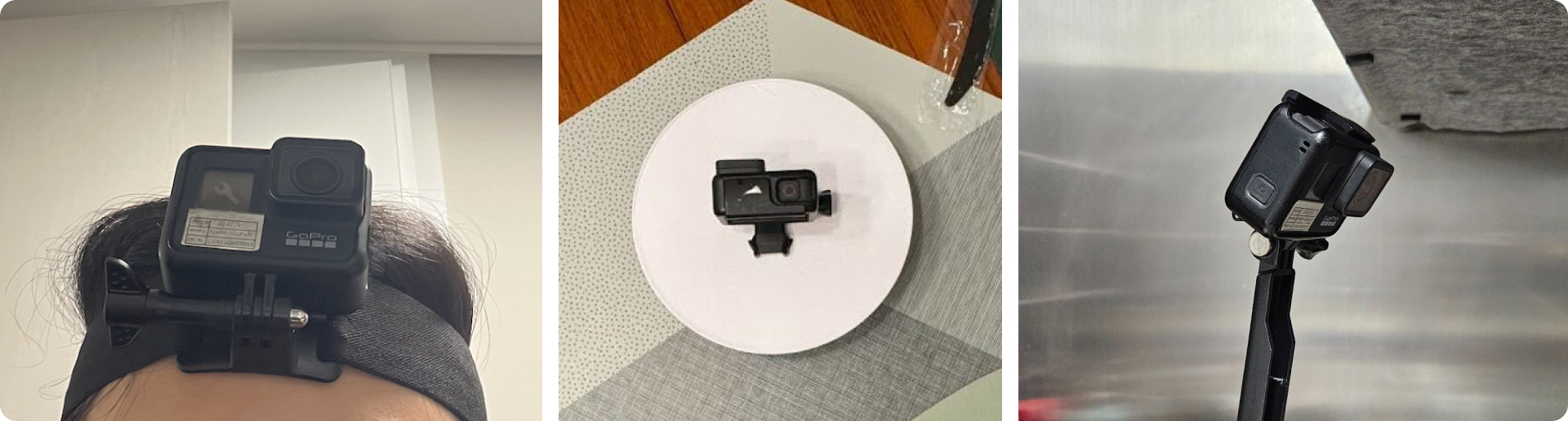}
    \caption{Examples of cameras and mounts used by participants (From left: each camera setup for simulating smart glasses, a 360-degree rotating camera, and a fixed kitchen camera)}
    \label{fig:enter-label}
    \Description{Wearing a head-mounted camera for smart glasses, using a rotating platform for a 360-degree camera, and securing the camera with a fixed mount.}
\end{figure}

\subsubsection{Diary-Style DIY Toolkit Concept Design}

The diary provided to participants was designed to systematically guide them through the key elements to consider at each step of simulating a DIY smart home setup using VLM camera sensors. To achieve this, we first reviewed a prior study \cite{yun2023potential} that focuses on the utilization of sensors in the DIY smart home construction process. This study outlines a five-step DIY smart home sensor utilization process, highlighting the critical factors that users must consider at each step. While incorporating the factors proposed in the prior study, we also tailored the diary to better align with the VLM camera sensor context. The diary we created follows the five-step DIY process suggested in the prior study. These five steps can be briefly summarized as follows: After selecting the feature to be developed, in Step 1, participants choose the situations that need to be sensed and the necessary sensors. In Step 2, they determine the roles of the camera sensors to effectively detect the selected situations. Step 3 involves selecting the optimal placement for these sensors. In Step 4, participants create feature rules. Finally, in Step 5, they test the feature and make adjustments. Detailed guides were provided to participants for each step, and we will explain these guides in more detail below. To support the explanation, we will use the example of P3’s DIY process to develop a "feature that provides helpful information when retrieving food from the refrigerator." 

\textbf{Step 1. Selecting sensing situation and sensor:} In this step, participants first determine the situation they want to sense, followed by selecting the appropriate camera sensor to detect that situation. Participants chose a smart home feature they wished to build through DIY and determined the situation that needed to be detected to implement that feature. For example, P3 decided to build a feature that “provides helpful information when retrieving food from the refrigerator”, to help prevent herself and her husband from consuming unhealthy foods and encourage healthier cooking. Therefore, she determined that it was necessary to detect which food item was being taken out of the refrigerator. After defining the sensing situation, they had to select a sensor type suitable for implementing this feature. P3 chose an internal refrigerator camera. Figure 3 shows the Step 1 section of the diary written by P3.

\begin{figure}
    \centering
    \includegraphics[width=1\linewidth]{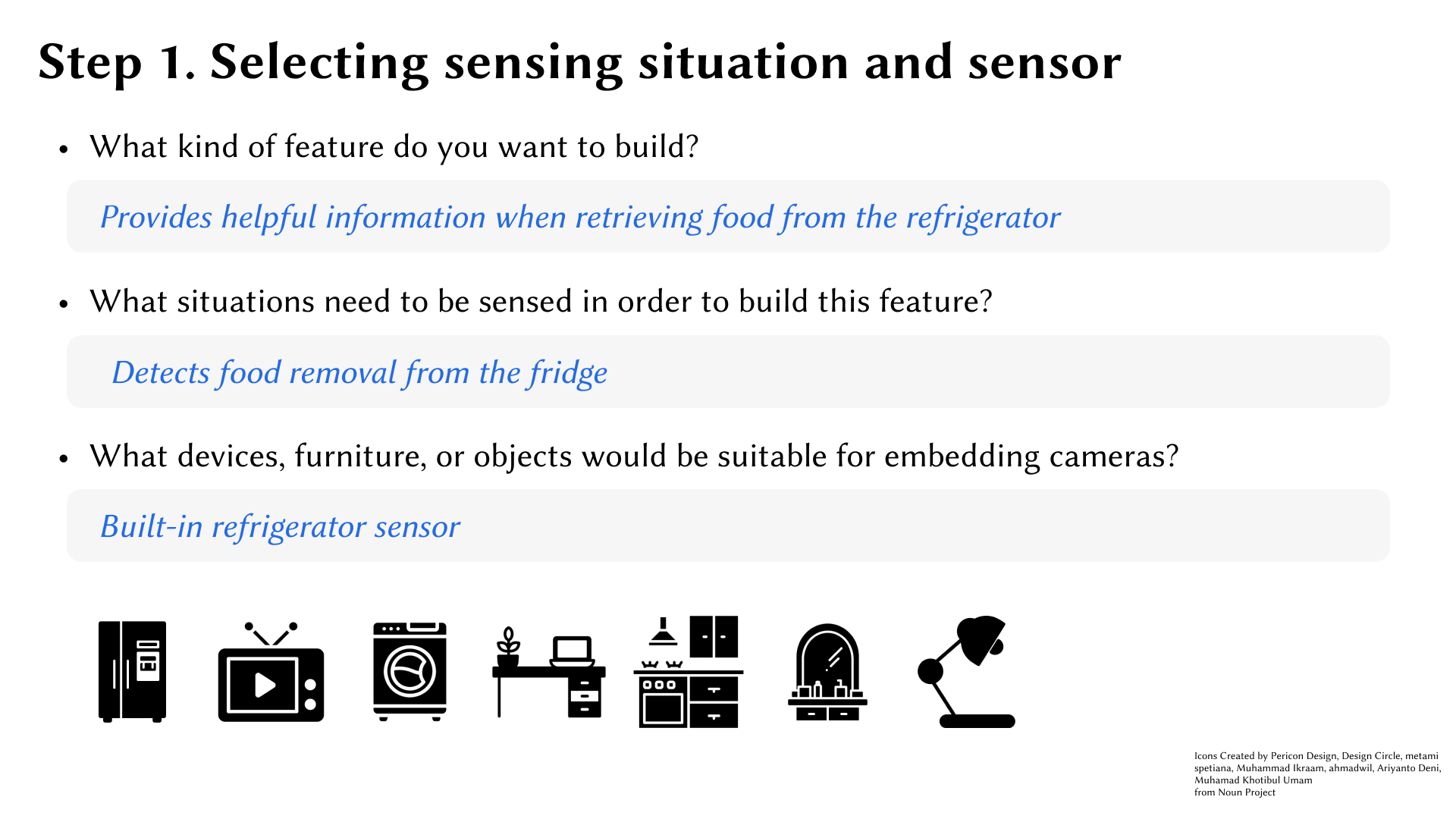}
    \caption{Step 1 of P3's Diary}
    \label{fig:enter-label}
    \Description{The diary includes the following questions: What kind of feature do you want to build?, What situations need to be sensed in order to build this feature?, What devices, furniture, or objects would be suitable for embedding cameras?}
\end{figure}

\textbf{Step 2. Determining the specific sensor role:} In this step, participants were asked to define (1) the role of the camera sensor and (2) the format of the output sensor data. First, to simulate the selected sensor type from Step 1, participants were instructed to position the camera using appropriate equipment and record video footage. Accordingly, P3 attached a camera inside the refrigerator and recorded a video. After recording, participants were asked to review the footage and capture key images (Figure 4-(a)). Next, participants specified the purpose of the image analysis by defining the sensor’s role. For example, after identifying the food item taken out of the refrigerator, P3 requested that the sensor analyze whether (a) the food was suitable to eat at that time for someone on a diet, (b) the appropriate portion size, and (c) which recipes would be healthy options. Additionally, participants were guided to specify the output format of the sensor data to ensure they received results in the desired format. P3 requested the results to be displayed in the following format (Figure 4-(b)): “ \_\_\_ is (not) suitable for consumption at the current time. The recommended portion size is \_\_\_. To enjoy this food healthily, \_\_\_.” Finally, participants uploaded the captured images, along with the sensor’s role and the desired output format, to ChatGPT using a temporary chat mode to protect their image data. They then tested whether the system could correctly analyze the images and generate sensor outputs as specified (Figure 5), and participants attached the test results to their diaries (Figure 6).

\begin{figure}%
\subfloat[]{{\includegraphics[width=0.47\textwidth ]{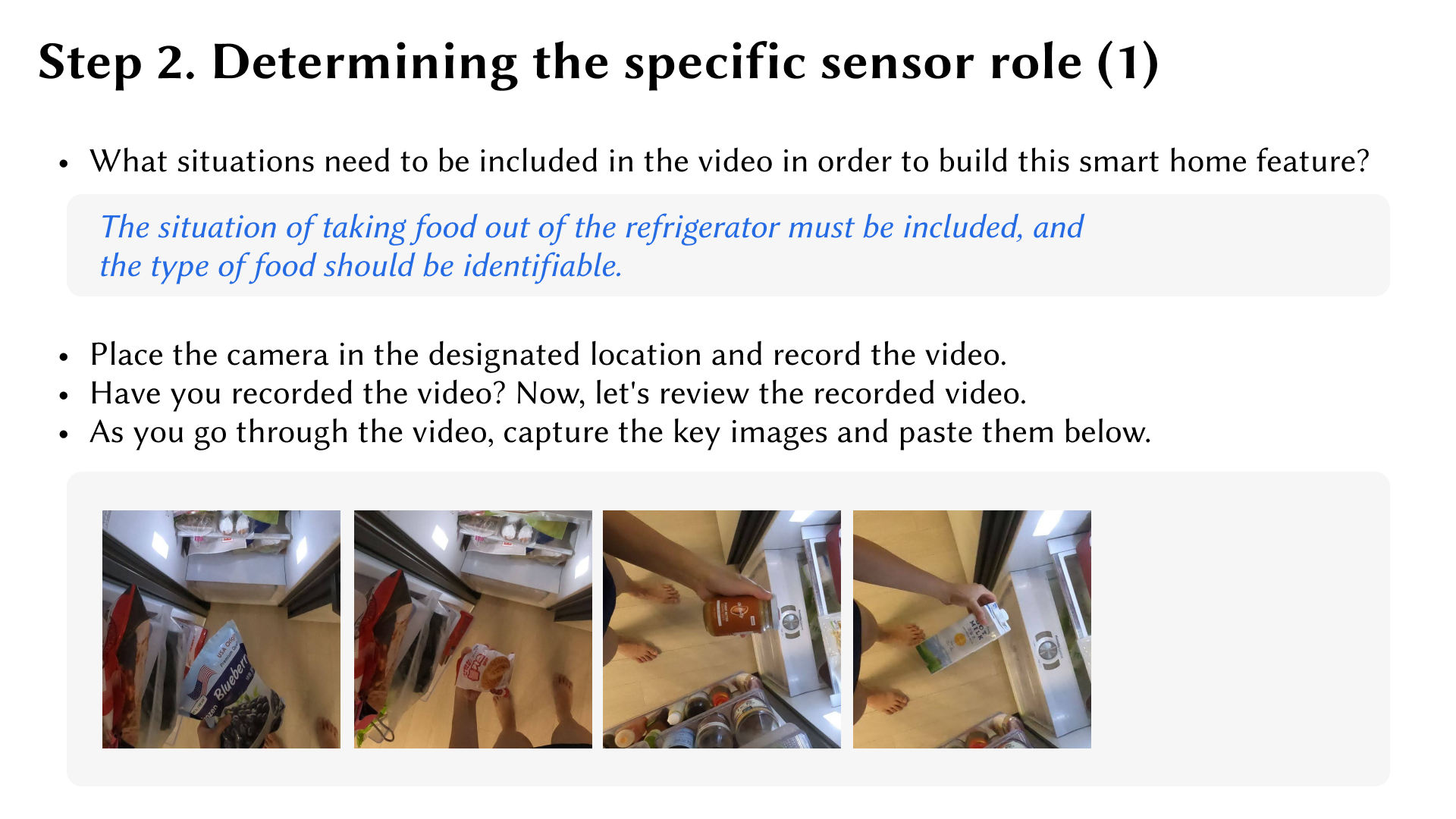}}}%
\vfill
\subfloat[]{{\includegraphics[width=0.47\textwidth ]{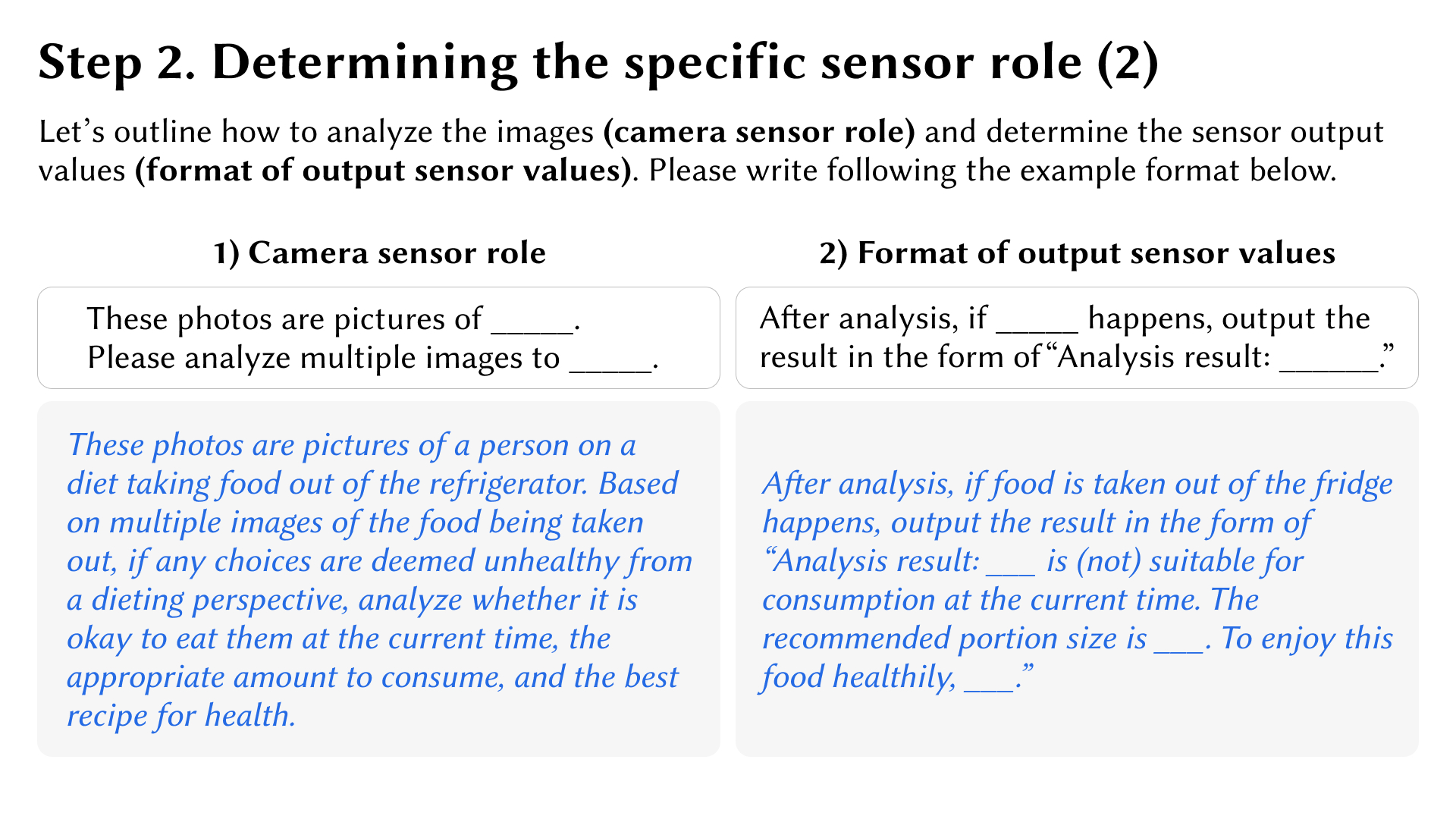}}}%
\caption{Step 2 of P3's Diary, (a) capturing the necessary footage and key images, (b) determining the role of the camera sensor and the format of the output sensor data}%
\label{tofsignal}%
\end{figure}

\begin{figure}
    \centering
    \includegraphics[width=1\linewidth]{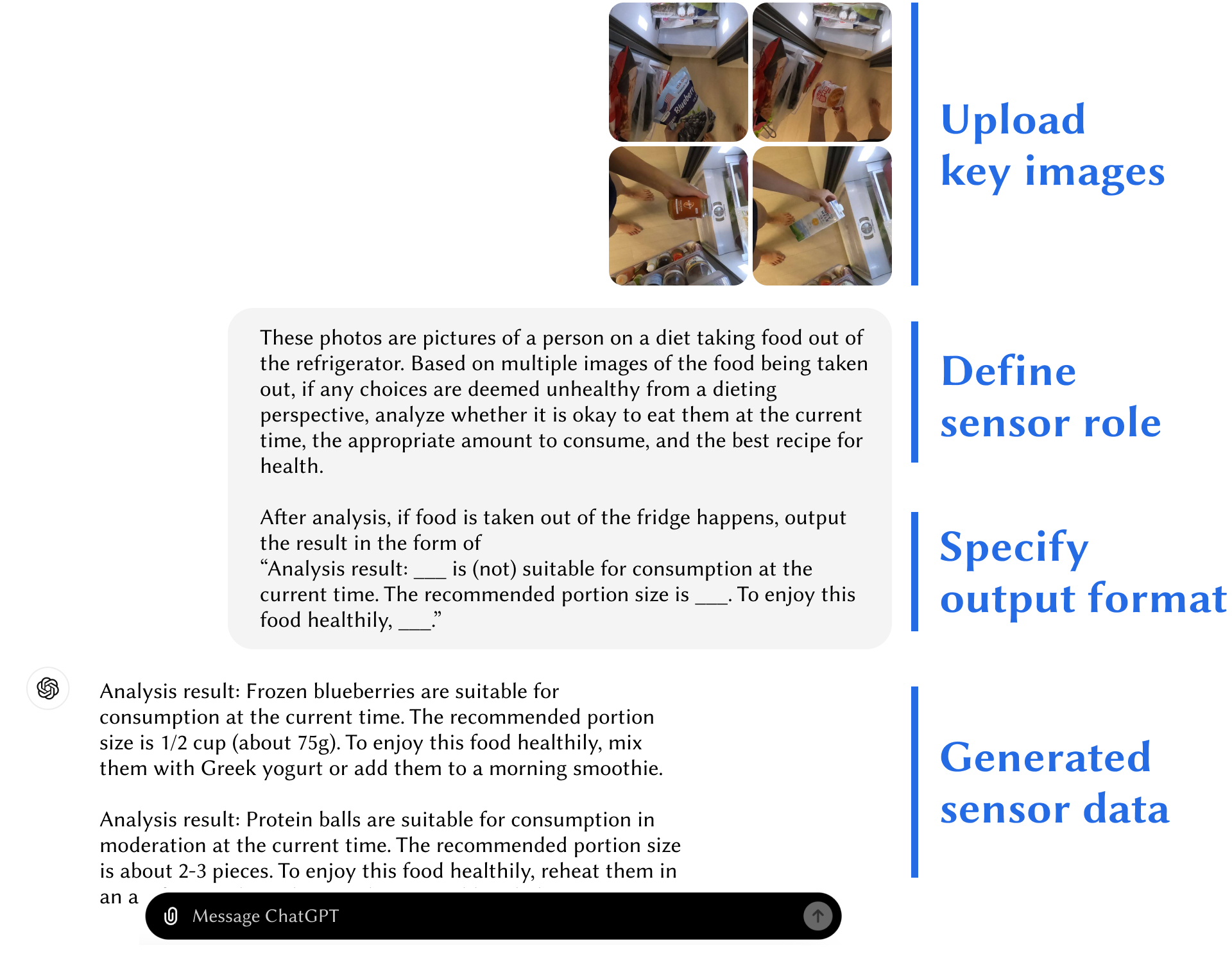}
    \caption{Example photo of P3 utilizing ChatGPT to receive sensor data based on captured images and a prompt specifying the camera sensor’s role and the format of the output sensor data.}
    \label{fig:enter-label}
\end{figure}

\begin{figure}
    \centering
    \includegraphics[width=1\linewidth]{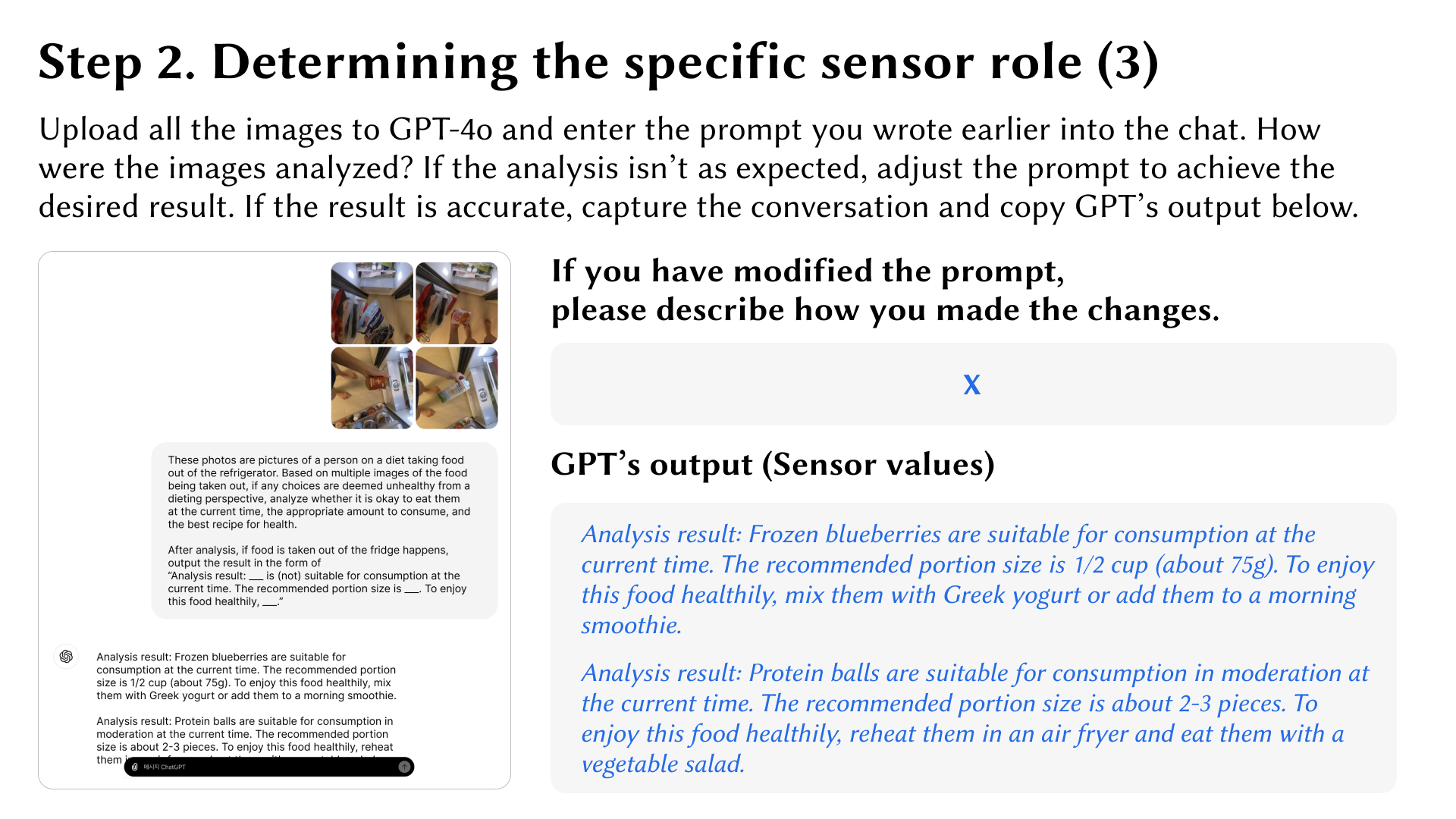}
    \caption{Step 2 of P3's Diary, testing with GPT and attaching the results}
    \Description{The diary includes the following questions: Where would be the best location to place the camera to perform the sensing role?, Why did you choose that location?, Place the camera in the selected locations, take pictures of how they are set up, and upload the images}
    \label{fig:enter-label}
\end{figure}

\textbf{Step 3. Placing the sensor appropriately:} In this step, participants determined the optimal location for the sensor to effectively perform its role. They tested filming from various angles to find the best position for capturing the desired elements. P3 decided that placing the camera above the refrigerator would best capture which food items were being retrieved, and used an alligator clip mount to secure the camera in that position (Figure 7).

\begin{figure}
    \centering
    \includegraphics[width=1\linewidth]{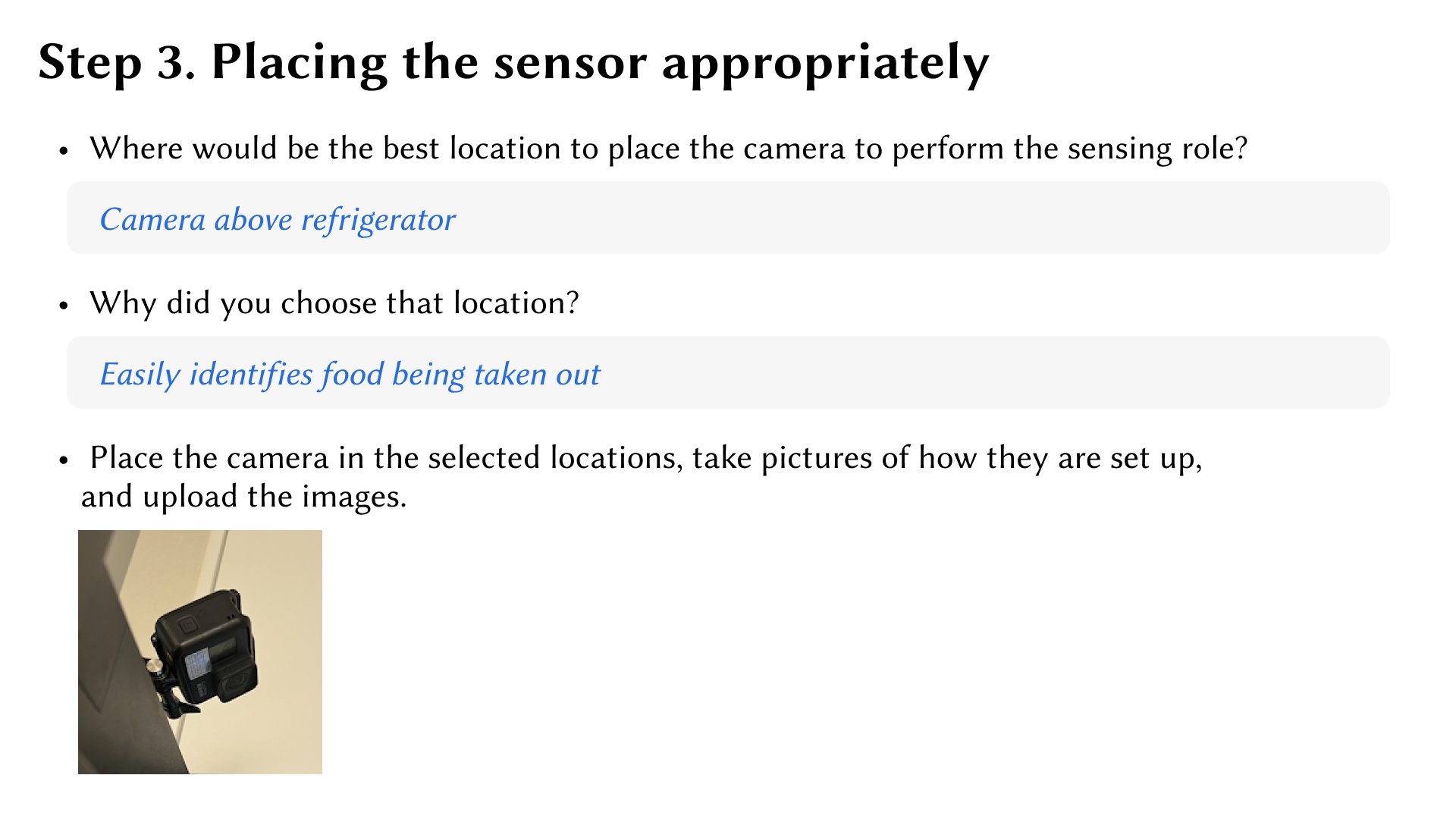}
    \caption{Step 3 of P3's Diary}
    \label{fig:enter-label}
\end{figure}

\textbf{Step 4. Making rules as intended:} This step involved setting up feature rules based on the sensor’s detected values. Participants used the If-this-then-that \cite{huang2015supporting, yu2021analysis} format to configure the smart home features. P3 defined rules to trigger different actions depending on whether the retrieved food was suitable for their diet. For example, P3 configured the system to provide positive feedback when diet-friendly food was selected, and to display a warning message for non-diet-friendly items (Figure 8-(a)). If it might not be necessary for the home camera to analyze daily activities continuously \cite{pierce2022addressing}, participants were asked to create rules ensuring the camera sensor operated only when needed. In P3’s case, the camera was set to turn on and start analyzing only when the refrigerator door was opened (Figure 8-(b)). While the camera couldn’t actually turn on/off based on these rules during our study, in Step 5, we asked participants to manually start and stop filming following the rules they had set.

\begin{figure}%
\subfloat[]{{\includegraphics[width=0.47\textwidth ]{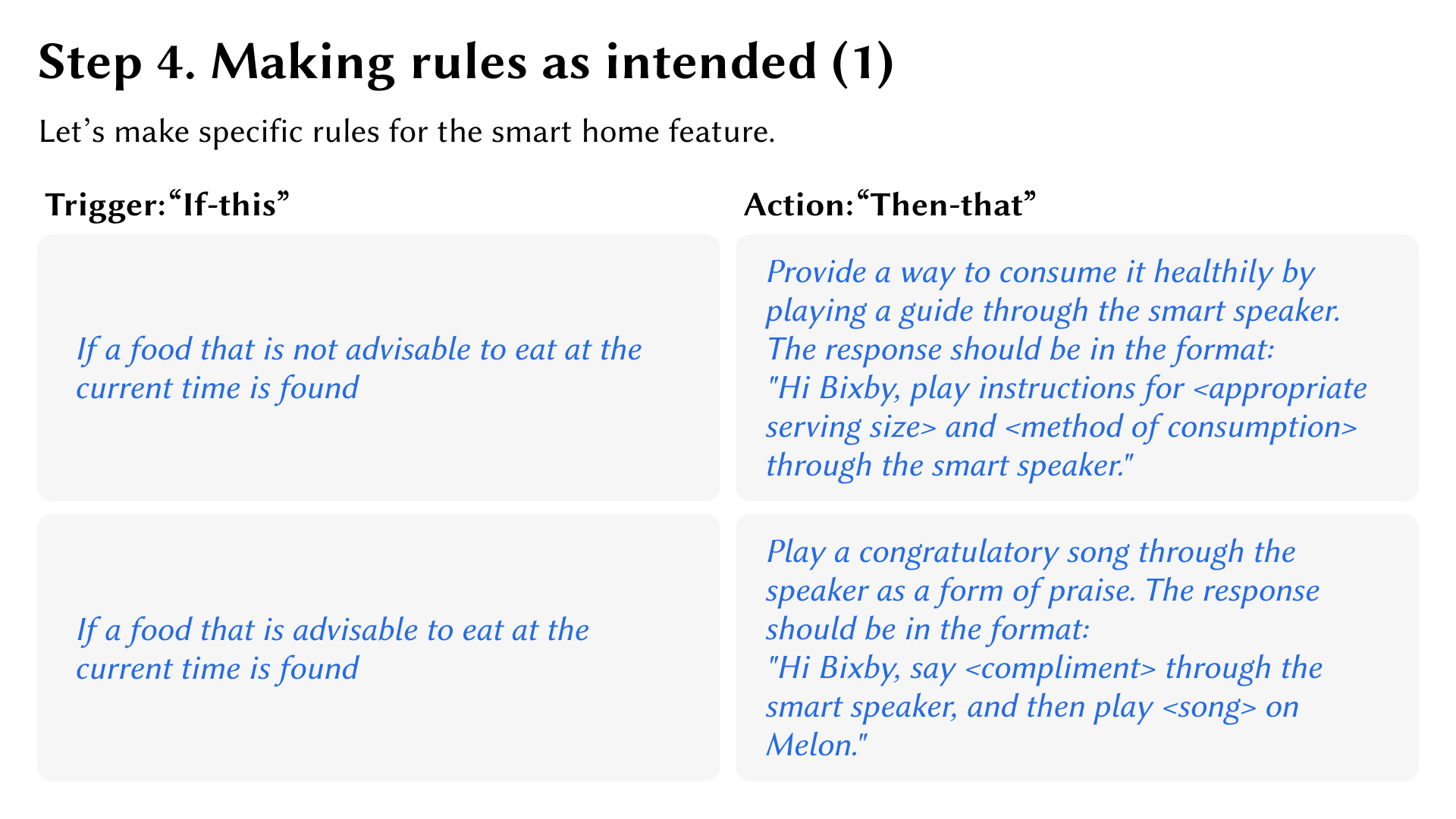}}}%
\hfill
\subfloat[]{{\includegraphics[width=0.47\textwidth ]{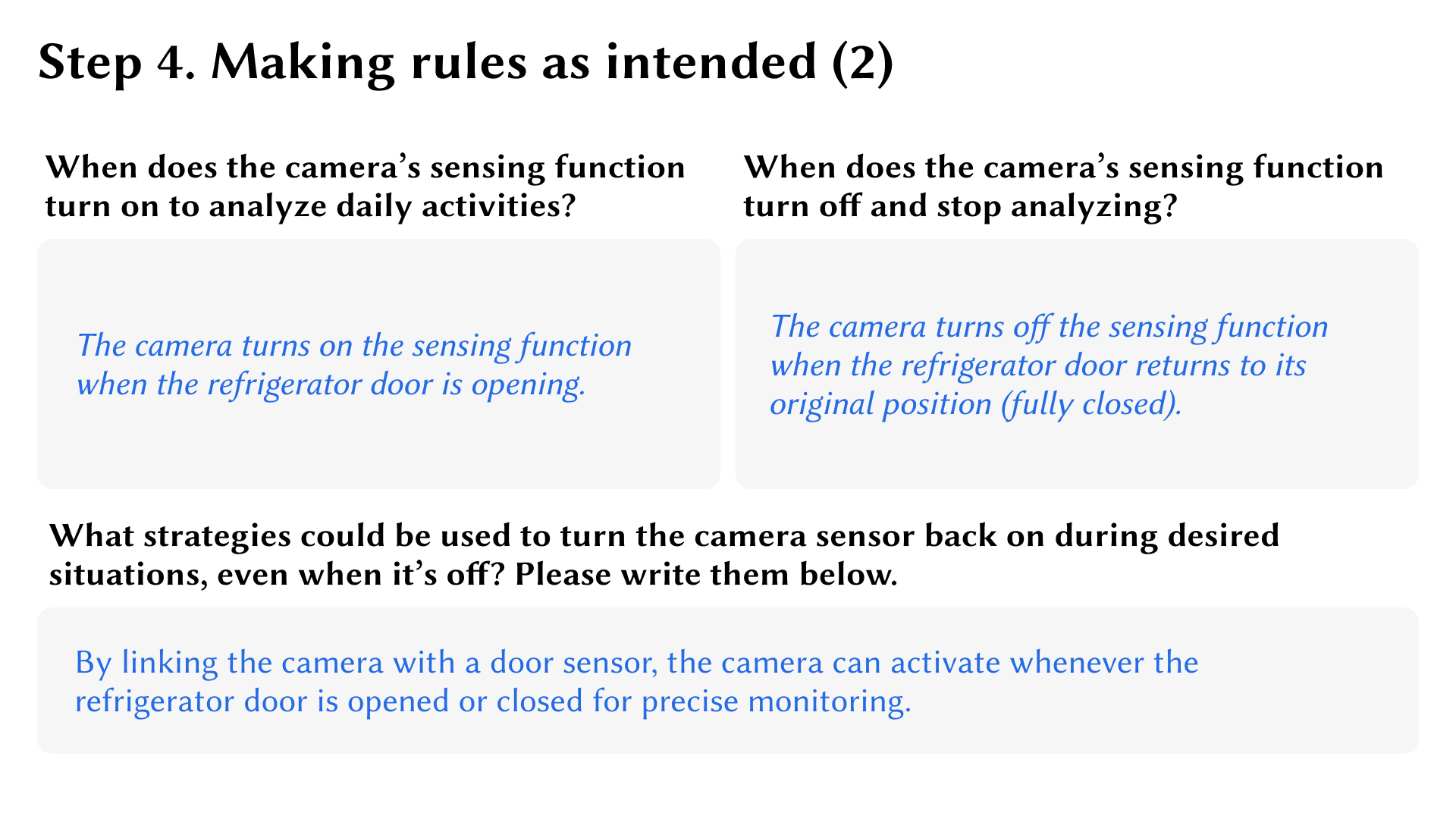}}}%
\caption{Step 4 of P3's Diary, (a) defining smart home feature rules using the If-this-then-that format, (b) setting camera sensor operation rules}%
\end{figure}

\textbf{Step 5. Testing and modifying feature:} In this step, participants tested whether the feature worked as intended and made adjustments if necessary. After installing the camera in the location selected during Step 3, they were instructed to record footage on the “camera sensor operation rules” defined in Step 4. Next, they combined the “sensor role,” “output data format” from Step 2, and the “rules for the feature” from Step 4 into a final prompt. Once the final prompt was completed, participants uploaded the captured images with the prompt to ChatGPT and tested the feature (Figure 9). To help participants better envision how the feature would work in real life, ChatGPT’s responses were framed as commands in the style of “Hi Bixby/Siri/OK Google, \_\_.” If the test results did not meet expectations, participants revised the prompt and continued testing iteratively until the desired outcome was achieved. Participants attached the generated results to their diaries (Figure 10).

\begin{figure}
    \centering
    \includegraphics[width=1\linewidth]{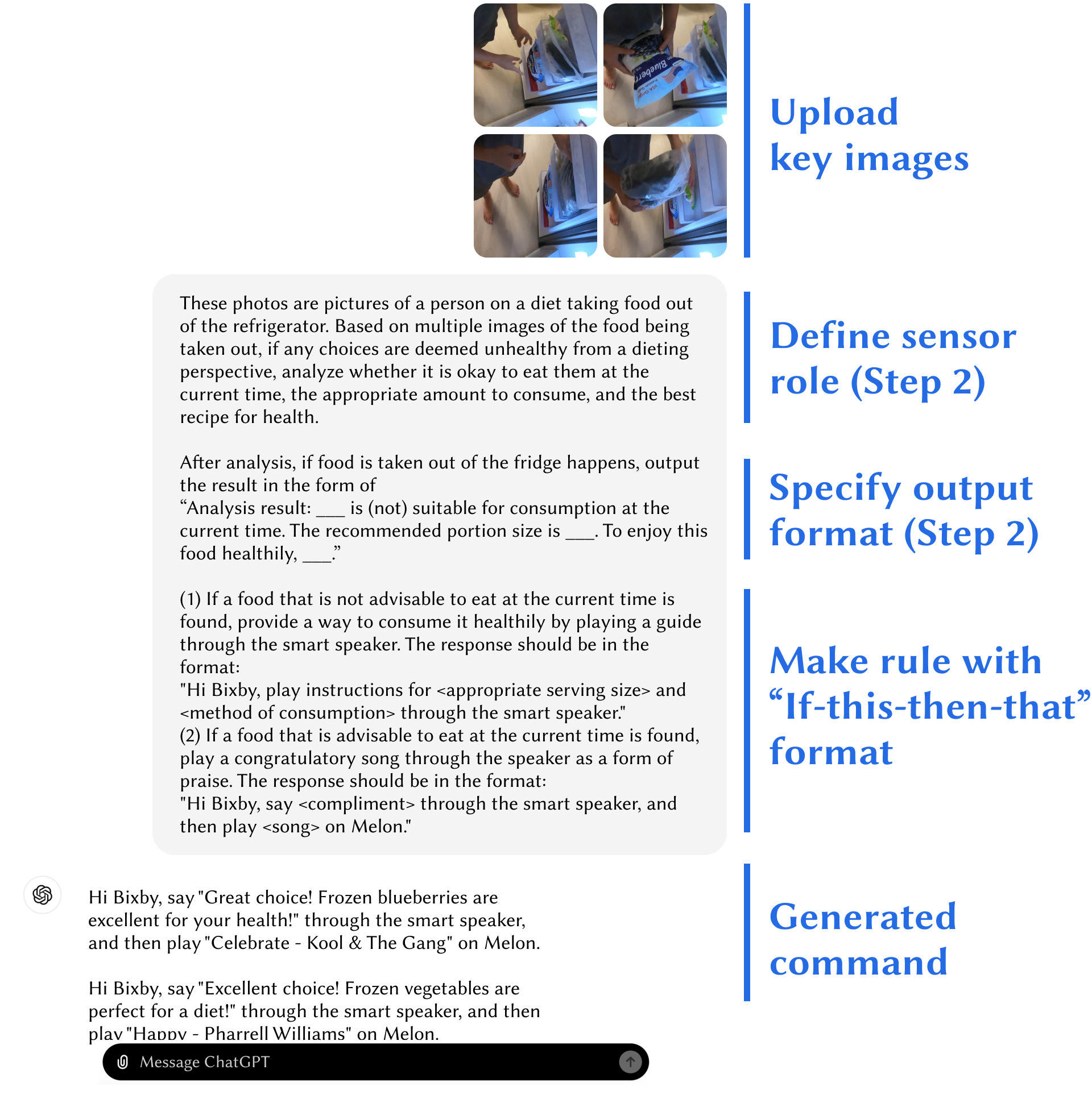}
    \caption{P3’s example of the generated results based on the captured images and the final prompt.}
    \label{fig:enter-label}
\end{figure}

\begin{figure}
    \centering
    \includegraphics[width=1\linewidth]{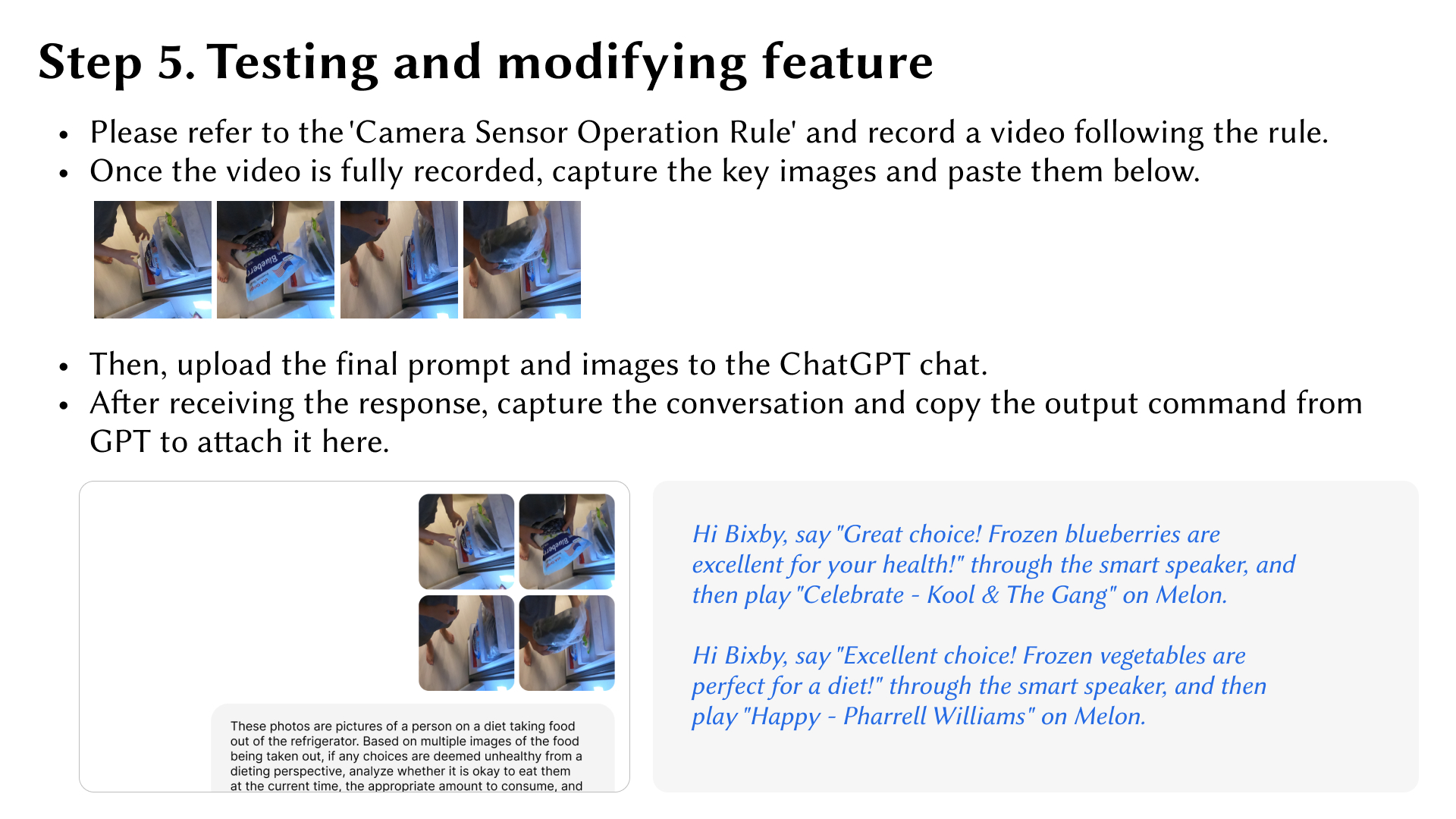}
    \caption{Step 5 of P3's Diary}
    \label{fig:enter-label}
\end{figure}

\subsection{Participants}

To gather a wide range of needs, we recruited 12 households with diverse family types and asked each household to select one representative (Table 1). The study aimed to explore the user experience of VLM camera sensor-based DIY smart homes compared to traditional DIY smart home experiences. Therefore, we recruited participants who had at least some experience in setting up smart home systems. Additionally, we ensured diversity in terms of gender, age, and smart home setup experiences. All participants attended a pre-study briefing session, and only those who agreed to participate were included in the study. This study received IRB approval (IRB No.KH2024-052).

\begin{table*}[]
\caption{Information of 12 participants involved in this study}
\label{tab:freq}
\begin{tabular}{cccccc}
\toprule[1pt]
\textbf{\#} & \textbf{Family type} & \textbf{Age} & \textbf{Gender} & \textbf{\begin{tabular}[c]{@{}c@{}}Duration\\ of setup\end{tabular}} & \textbf{\begin{tabular}[c]{@{}c@{}}Participants’ existing\\ DIY smart home configurations\end{tabular}} \\ \midrule[1pt]
P1 & \begin{tabular}[c]{@{}c@{}}4-person household,\\ living in an apartment\\ \end{tabular} & 20 & Male & 6 years & \begin{tabular}[c]{@{}c@{}}Uses SmartThings and Aqara hub for smart lighting, \\ temperature control, motorized curtains, etc.\\ \end{tabular} \\ \midrule[0.5pt]
P2 & \begin{tabular}[c]{@{}c@{}}4-person household,\\ living in an apartment\end{tabular} & 39 & Male & 2.5 years & \begin{tabular}[c]{@{}c@{}}SmartThings for blind adjustments,\\ lighting controls, and boiler automation.\end{tabular} \\ \midrule[0.5pt]
P3 & \begin{tabular}[c]{@{}c@{}}2-person household,\\ living in an apartment\end{tabular} & 28 & Female & 1.5 years & \begin{tabular}[c]{@{}c@{}}Primarily uses SmartThings and Google Assistant \\ for turning off lights and checking gas locks before bed.\\ \end{tabular} \\ \midrule[0.5pt]
P4 & \begin{tabular}[c]{@{}c@{}}1-person household,\\ living in an apartment\end{tabular} & 25 & Female & - & \begin{tabular}[c]{@{}c@{}}Owns smart lighting and sensors but hasn't fully\\ implemented due to the setup effort required.\\ \end{tabular} \\ \midrule[0.5pt]
P5 & \begin{tabular}[c]{@{}c@{}}3-person household,\\ living in an apartment\end{tabular} & 34 & Female & 3 years & \begin{tabular}[c]{@{}c@{}}Uses a built-in smart home system and SmartThings sensors \\ for lighting, humidity, CO2, and leak detection.\\ \end{tabular} \\ \midrule[0.5pt]
P6 & \begin{tabular}[c]{@{}c@{}}3-person household,\\ living in a detached house\\ \end{tabular} & 39 & Male & 8 years & \begin{tabular}[c]{@{}c@{}}Integrates SmartThings and Home Assistant (HA) for lighting,\\ CCTV detection, baby cry alerts, and indoor temperature control.\\ \end{tabular} \\ \midrule[0.5pt]
P7 & \begin{tabular}[c]{@{}c@{}}1-person household,\\ living in an apartment\\ (sometimes with a cat)\end{tabular} & 29 & Male & 5 years & \begin{tabular}[c]{@{}c@{}}Uses a built-in smart home system for lighting and ventilation,\\ Apple Home to automatically turn on lights\\ and play music upon returning home.\end{tabular} \\ \midrule[0.5pt]
P8 & \begin{tabular}[c]{@{}c@{}}1-person household,\\ living in an apartment\end{tabular} & 25 & Male & 0.5 years & \begin{tabular}[c]{@{}c@{}}Uses TAPO bulbs for automatic lighting after bedtime\\ and has previous experience with HA, ThingQ, and SmartThings.\\ \end{tabular} \\ \midrule[0.5pt]
P9 & \begin{tabular}[c]{@{}c@{}}3-person household,\\ living in an apartment\end{tabular} & 23 & Female & - & \begin{tabular}[c]{@{}c@{}}Leverages SmartThings for adjusting air conditioner\\  and fridge features but lacks a complete automation setup.\\ \end{tabular} \\ \midrule[0.5pt]
P10 & \begin{tabular}[c]{@{}c@{}}1-person household,\\ studio apartment\end{tabular} & 25 & Female & - & \begin{tabular}[c]{@{}c@{}}Has experience building smart home\\ features but not in a complete setup.\end{tabular} \\ \midrule[0.5pt]
P11 & \begin{tabular}[c]{@{}c@{}}2-person household,\\ living in an apartment\\ (with a dog)\end{tabular} & 46 & Male & 6 years & \begin{tabular}[c]{@{}c@{}}Incorporates SmartThings and HA\\ to automate lighting\\based on human presence.\\ \end{tabular} \\ \midrule[0.5pt]
P12 & \begin{tabular}[c]{@{}c@{}}2-person household,\\ living in an apartment\end{tabular} & 32 & Male & 4 years & \begin{tabular}[c]{@{}c@{}}Built an HA-based automation system for away/home mode,\\ dining light control, and HVAC adjustment based on window status.\\ \end{tabular} \\ \bottomrule[1pt]
\end{tabular}
\end{table*}

\subsection{Study Procedure}

This study was conducted over three weeks, during which participants simulated building, testing, and refining DIY smart home features in their households using the online diary described in Section 3.1.2. To encourage a variety of ideas, we proposed different camera sensor configurations for smart home construction each week. Based on the concepts outlined in Section 3.1.1, the study progressed with the following themes: in Week 1, “a camera that follows and continuously observes daily life”; in Week 2, “a camera that synthesizes the overall situation at home”; and in Week 3, “a camera embedded in a device or object to focus on specific spaces.”

Each week began with an ice-breaking activity on the first day to help participants become familiar with the concept of the camera. From days 2 to 7, they followed the five-step diary process (Section 3.1.2) to implement 2–3 smart home features aligned with the weekly theme. The overall schedule is shown in Figure 11.

\begin{figure}
    \centering
    \includegraphics[width=1\linewidth]{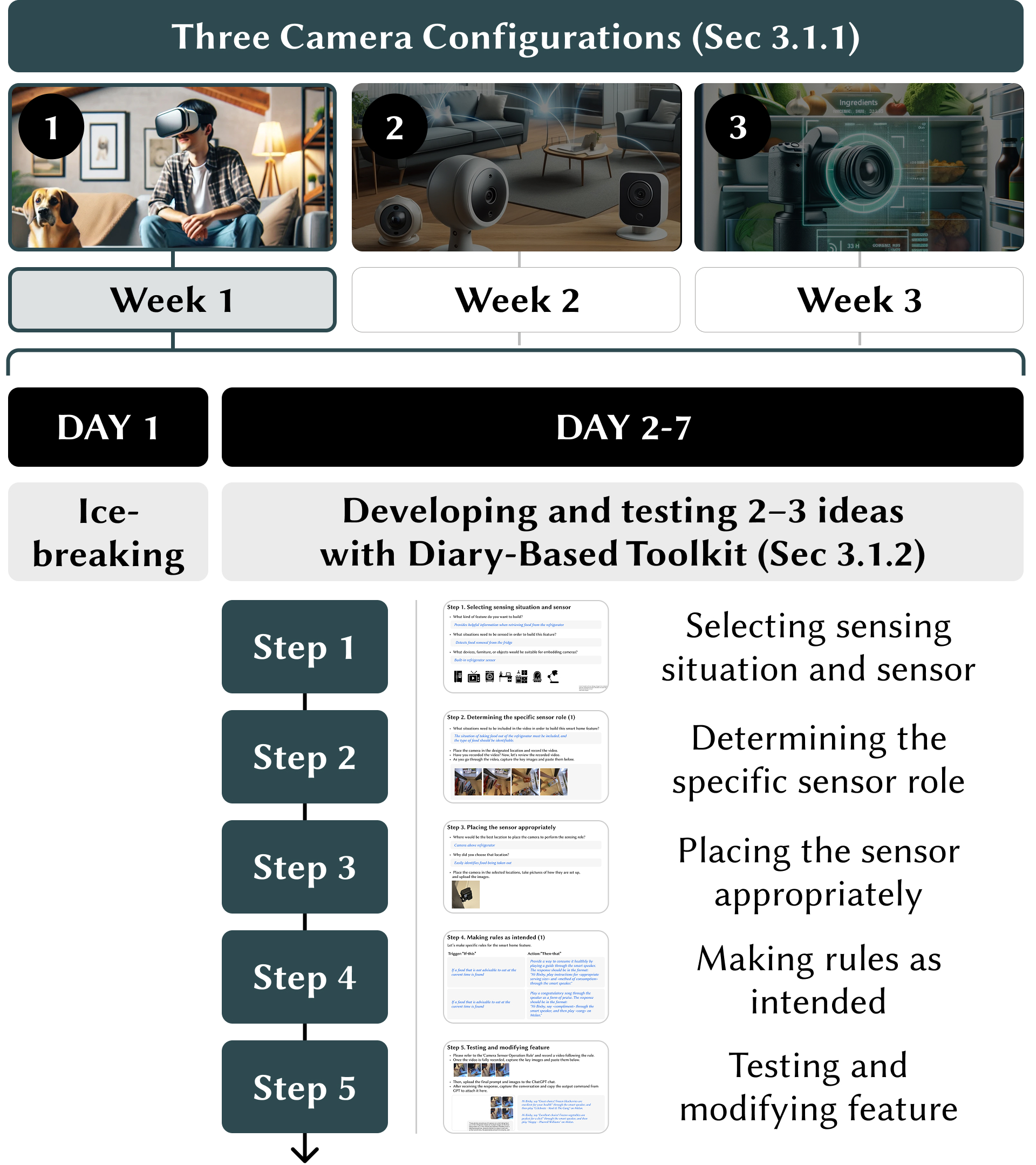}
    \caption{\textcolor{black}{Overview of the study schedule}}
    \Description{Each week began with an icebreaking activity on the first day to help participants become familiar with the camera concept. From days 2 to 7, they followed the five-step diary process (Section 3.1.2) to implement 2–3 smart home features aligned with the weekly theme.}
    \label{fig:enter-label}
\end{figure}

On the first day of each week, ice-breaking activities were conducted to help participants familiarize themselves with image analysis. Participants filmed videos based on missions provided by the researcher and analyzed the images using ChatGPT. These missions were designed around nine smart home feature categories from a previous study \cite{garg2022social}, which included enhancing family bonds, clothing management, meal preparation, relaxation, maintaining cleanliness and household items, pet care, health goals, recording schedules and memories, and personal goals. From the second to the seventh day of each week, participants developed and tested 2-3 ideas aligned with that week’s camera sensor concept. They were asked to complete these activities at their convenience over the six days. After completing the three-week session, each participant participated in a 90 to 120-minute interview. The interview was divided into three parts: 1) questions about their existing smart home experiences, 2) questions on the feature-building process with VLM camera sensors, and 3) questions about the potential and concerns of VLM camera sensor-based smart homes.

\subsection{Data Analysis}

As a result of the study, we collected 107 smart home feature ideas and dairies. The diaries served as key material in formulating interview questions. Before the interviews, we prepared semi-structured questions to compare participants' experiences of using traditional IoT sensors and VLM camera sensors in the process of building smart home features. The interviews were audio-recorded, collecting a total of approximately 1,257 minutes of interview data. All interview recordings were transcribed, and key quotes were extracted. We then performed qualitative thematic analysis on the interview data \citep{boyatzis1998transforming}, with two HCI researchers experienced in qualitative data analysis conducting open coding together \cite{urquhart2013using}.

Our analysis was primarily centered around the participants’ quotes, and to gain a more detailed understanding of the meanings behind these quotes, we referred to the smart home ideas participants derived and the diaries they wrote. The quotes were categorized according to three key themes we aimed to explore: 1) anticipated roles of VLM camera sensor-based smart homes, 2) considerations during the DIY development process, and 3) participants’ concerns. The categorized quotes were then further analyzed through thematic analysis to identify major themes. Based on this analysis, the following section will explain in detail the user needs for VLM camera sensor-based smart homes (Section 4.1), the characteristics and considerations during the development process (Section 4.2), and participants’ concerns (Section 4.3).

\section{Findings}

\subsection{What Roles Did Participants Expect from VLM Camera Sensor-based DIY Smart Homes?}

This section explains the roles that participants envisioned for DIY smart home features using VLM camera sensors, incorporating the three camera configuration concepts. Participants developed features in three primary roles: auto-monitoring (Section 4.1.1), assistant (Section 4.1.2), and advisory (Section 4.1.3). For a detailed list of the features created by participants, please refer to Appendix (Table 5). 

\subsubsection{Auto-Monitoring Role}

\begin{table*}[]
\caption{Examples of participant ideas for auto-monitoring role features}
\label{tab:freq}
\begin{tabular}{cccc}
\toprule[1pt]
\textbf{Developed Feature} & \textbf{Example Situation} & \textbf{If this}& \textbf{Then that}\\ 
\midrule[1pt]
\vspace{0.4mm}
\begin{tabular}[c]{@{}c@{}}Alerts when electricity\\ is being wasted\\ unnecessarily (P8)\end{tabular} & \begin{minipage}[c]{.25\textwidth}
      \includegraphics[width=\textwidth]{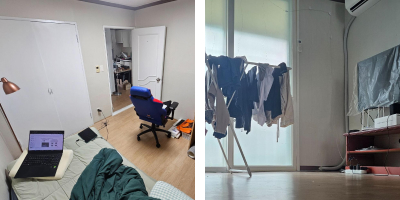}
    \end{minipage}& \begin{tabular}[c]{@{}c@{}}If unnecessary\\ appliances\\ are running\end{tabular} & \begin{tabular}[c]{@{}c@{}}Turn off\\ unnecessary\\ appliances\end{tabular} \\ \midrule[0.5pt]\vspace{0.4mm}
\begin{tabular}[c]{@{}c@{}}Alerts for\\ dangerous situations\\ in the kitchen (P1)\end{tabular} & \begin{minipage}[c]{.25\textwidth}       \includegraphics[width=\textwidth]{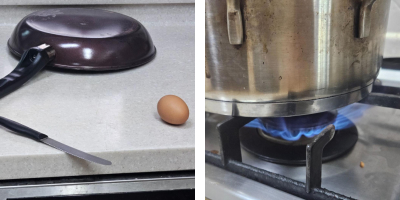}     \end{minipage} & \begin{tabular}[c]{@{}c@{}}If a potentially dangerous\\ situation occurs\\ while cooking\end{tabular} & \begin{tabular}[c]{@{}c@{}}Alert about\\ the dangerous\\ situation\end{tabular} \\ \midrule[0.5pt]\vspace{0.4mm}
\begin{tabular}[c]{@{}c@{}}Automatic pet care\\ for when pets\\ need help (P11)\end{tabular} & \begin{minipage}[c]{.25\textwidth}       \includegraphics[width=\textwidth]{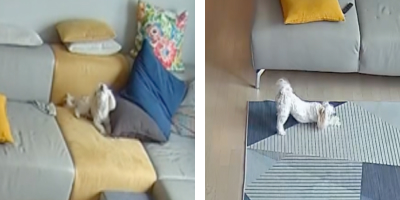}     \end{minipage} & \begin{tabular}[c]{@{}c@{}}If the dog shows \\ separation anxiety\\ while alone at home\end{tabular} & \begin{tabular}[c]{@{}c@{}}Play white noise\\ to reduce anxiety\end{tabular} \\ \midrule[0.5pt]\vspace{0.4mm}
\begin{tabular}[c]{@{}c@{}}Alerts when\\ posture needs\\ correction (P4)\end{tabular} & \begin{minipage}[c]{.25\textwidth}       \includegraphics[width=\textwidth]{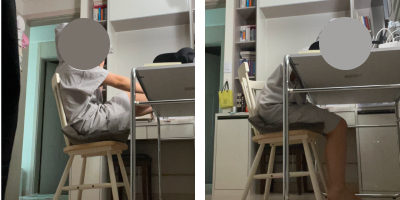}     \end{minipage} & \begin{tabular}[c]{@{}c@{}}If I am working\\ and need\\ posture correction\end{tabular} & \begin{tabular}[c]{@{}c@{}}Play a voice prompt\\ to remind me\\ to correct my posture\end{tabular} \\ \midrule[0.5pt]\vspace{0.4mm}
\begin{tabular}[c]{@{}c@{}}Turns on the light\\ when the child is \\ doing activities without\\ sufficient lighting (P2)\end{tabular} & \begin{minipage}[c]{.25\textwidth}       \includegraphics[width=\textwidth]{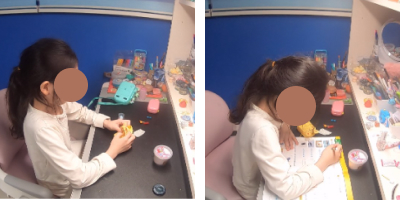}     \end{minipage} & \begin{tabular}[c]{@{}c@{}}If the child is doing\\ an activity at the desk\\ without sufficient lighting\end{tabular} & \begin{tabular}[c]{@{}c@{}}Adjust lighting\\ and play\\ appropriate music\end{tabular} \\ 
\bottomrule[1pt] \vspace{0.4mm}
\end{tabular}
\end{table*}

Participants leveraged the wide field of view and situational analysis capabilities of VLM camera sensors to develop DIY smart home features for automatic home monitoring. These features tracked the home environment, user behavior, and family members' situations, triggering alerts or automated actions when necessary (Table 2). The goal was to ensure automatic monitoring of critical home situations requiring attention.

Participants built features that leveraged the camera sensor for auto-monitoring homes, allowing them to receive notifications about spaces that required management in terms of hygiene, tidiness, safety, and energy efficiency.

\begin{quote}\textit{
    Even if I’m not looking, if there’s an unsafe situation in the kitchen, like an egg about to fall, a knife slipping, or smoke starting, I thought of a feature where the camera could alert me. (P1)}
\end{quote}

Additionally, participants developed features to enhance awareness of their own and their family members' situations. By utilizing camera sensors to track their behavior, they identified unconscious habitual actions, fostering opportunities to adopt healthier lifestyle habits. These auto-monitoring features were especially useful for households with children or pets requiring constant care. Participants created features to detect moments when children or pets might need assistance, even in their absence, enabling the system to send notifications or trigger automatic responses. This approach supported more efficient caregiving tasks and promoted a safer home environment.
\begin{quote}\textit{    
    I’ve been trying to stop biting my nails for almost 10 years, but I still haven’t kicked the habit. Since it’s something I do subconsciously when I’m focused, I thought it would be great if the camera could detect my behavior and warn me. (P10)}
\end{quote}
\begin{quote}\textit{   
    I built a smart home feature that triggers commands like ‘No,’ ‘Stop,’ or ‘Wait’ in my voice when my dog chews on the couch or pees in the wrong place, so the situation can be addressed immediately. (P11)}
\end{quote}

Participants anticipated that auto-monitoring features would provide reassurance that no concerning situations would occur at home, even when they were away or occupied. They also viewed the ability to analyze daily life through the camera sensor as an opportunity to gain an objective perspective on their home. This perspective enabled them to recognize overlooked issues and reflect on familiar routines, making it a satisfying experience for them. 
\begin{quote}\textit{  
    If the smart home tells me my desk is messy, I can objectively see that, yes, it is messy. I think the camera could also bring attention to situations I subconsciously know but am avoiding. [...] It gave me great satisfaction to have the opportunity to step back and reflect on the familiar scenery of my home from a fresh perspective. (P10)} 
\end{quote}

In conclusion, participants developed features to automatically monitor home situations using VLM camera sensors, enabling alerts or automatic actions during critical moments that required attention to the state of the home, family members, or their own behavior. 

\subsubsection{Assistant Role}

\begin{table*}[]
\caption{Examples of participant ideas for assistant role features}
\label{tab:freq}
\begin{tabular}{cccc}
\toprule[1pt]
\textbf{Developed Feature} & \textbf{Example Situation} & \textbf{If this}& \textbf{Then that}\\ 
\midrule[1pt]
\vspace{0.4mm}
\begin{tabular}[c]{@{}l@{}}Guides organization\\ when items are not\\ arranged according\\ to the rules (P7)\end{tabular} & \begin{minipage}{.25\textwidth}       \includegraphics[width=\textwidth]{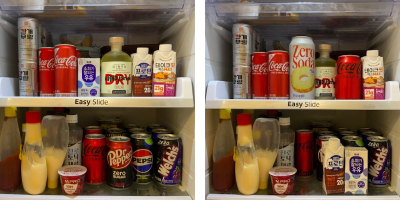}     \end{minipage} & \begin{tabular}[c]{@{}l@{}}If drinks of the\\ same type are placed\\ in different sections\end{tabular} & \begin{tabular}[c]{@{}l@{}}Prompt me\\ to sort and\\ organize the items\end{tabular} \\ \midrule[0.5pt]\vspace{0.4mm}
\begin{tabular}[c]{@{}l@{}}Sequentially guides\\ me through\\ a coffee recipe (P7)\end{tabular} & \begin{minipage}{.25\textwidth}       \includegraphics[width=\textwidth]{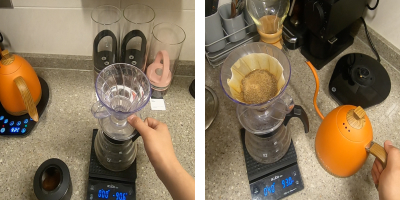}     \end{minipage} & \begin{tabular}[c]{@{}l@{}}If I start brewing coffee\\ using a hand drip,\\ prompt me to choose\\ which recipe to follow\end{tabular} & \begin{tabular}[c]{@{}l@{}}Provide step-by-step\\ guidance on\\ what to do next\end{tabular} \\ \midrule[0.5pt]\vspace{0.4mm}
\begin{tabular}[c]{@{}l@{}}Helps meticulously\\ check the condition\\ of clothes (P9)\end{tabular} & \begin{minipage}{.25\textwidth}       \includegraphics[width=\textwidth]{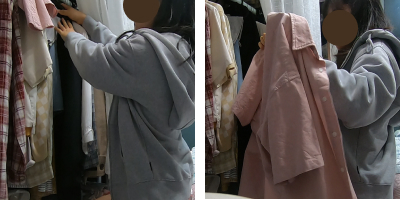}     \end{minipage} & \begin{tabular}[c]{@{}l@{}}If I'm organizing clothes\\ in front of the closet\end{tabular} & \begin{tabular}[c]{@{}l@{}}Recommend clothes\\ for the AirDresser\\ and activate\\ the appropriate mode\end{tabular} \\ \midrule[0.5pt]\vspace{0.4mm}
\begin{tabular}[c]{@{}l@{}}Assists with\\ studying (P8)\end{tabular} & \begin{minipage}{.25\textwidth}       \includegraphics[width=\textwidth]{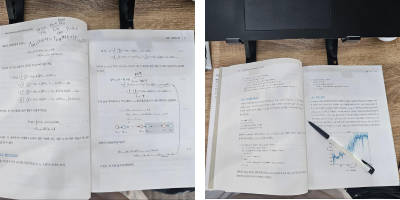}     \end{minipage} & \begin{tabular}[c]{@{}l@{}}If I open\\ a book to study\end{tabular} & \begin{tabular}[c]{@{}l@{}}Analyze the key points\\ of the page and\\ summarize the content\end{tabular} \\ \midrule[0.5pt]\vspace{0.4mm}
\begin{tabular}[c]{@{}l@{}}Assists with baby\\ care actions (P8)\end{tabular} & \begin{minipage}{.25\textwidth}       \includegraphics[width=\textwidth]{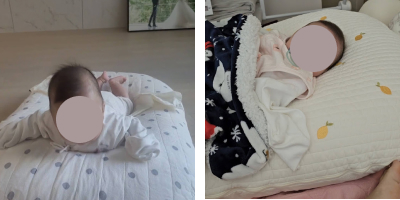}     \end{minipage} & \begin{tabular}[c]{@{}l@{}}When my nephew visits,\\ analyze the\\ baby's expression\end{tabular} & \begin{tabular}[c]{@{}l@{}}Analyze whether\\ appropriate caregiving\\ actions were taken\end{tabular} \\ \bottomrule[1pt]\vspace{0.4mm}
\end{tabular}
\end{table*}

Participants perceived VLM sensors as more meticulous and sensitive than themselves, with the ability to provide useful information about current situations. Leveraging these strengths, participants developed features that offered enriched support for their daily routines (Table 3).

Participants envisioned the camera sensor as an assistant that analyzes situations alongside them, especially in cases where human judgment might miss subtle details. This role was particularly useful for detecting subtle changes or reviewing large volumes of information. For those with limited knowledge of certain situations, the assistant feature supported decision-making, which was especially valuable for participants caring for infants or pets. For example, P8, who cared for their niece, and P7, who looked after their partner’s pet, developed features to monitor non-verbal behaviors, enabling them to assess the comfort of those under their care.
\begin{quote}\textit{
    Cameras can detect subtle changes like dry soil or drooping leaves better than a person. (P5)}
\end{quote}
\begin{quote}\textit{
    The camera can recognize subtle changes in the baby’s expression when I take a caregiving action. If the expression turns positive, it means I did the right thing, so I think it would be very effective in that sense. (P8)}
\end{quote}

Additionally, features that provided useful information about the current situation played a role in helping participants better understand and become more aware of their own circumstances.
\begin{quote}\textit{
    I’ve developed a habit of checking how much sugar is in food. I thought it would be really convenient if the camera could do that for me, so I built this feature. (P8)}
\end{quote}

Participants expressed satisfaction with the assistant features, as they enabled the effortless acquisition of useful information in daily life. This information supported the maintenance of an ideal routine and highlighted important but previously overlooked aspects of their everyday lives. 

In conclusion, participants developed assistant features that supported their activities and enriched their daily lives. These features offered new information and perspectives, making daily life more meaningful and fulfilling.

\subsubsection{Advisory Role}

\begin{table*}[]
    \caption{Examples of participant ideas for advisory role features}
    \label{tab:freq}
\begin{tabular}{cccc}
\toprule[1pt]
\textbf{Developed Feature} & \textbf{Example Situation} & \textbf{If this}& \textbf{Then that}\\ 
\midrule[1pt]
\vspace{0.4mm}
\begin{tabular}[c]{@{}l@{}}Checks the condition\\ of plants and\\ suggests solutions (P5)\end{tabular} & \begin{minipage}{.25\textwidth}       \includegraphics[width=\textwidth]{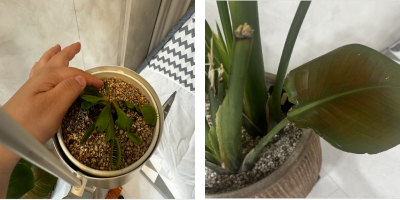}     \end{minipage} & \begin{tabular}[c]{@{}l@{}}1) If a change in\\ leaf color or damage\\ is detected\\ 2) If insects are found\\ on a specific plant\end{tabular} & \begin{tabular}[c]{@{}l@{}}Provide guidance\\ on how to address\\ the issue\end{tabular} \\ \midrule[0.5pt]\vspace{0.4mm}
\begin{tabular}[c]{@{}l@{}}Offers advice for\\ efficient tidying (P10)\end{tabular} & \begin{minipage}{.25\textwidth}       \includegraphics[width=\textwidth]{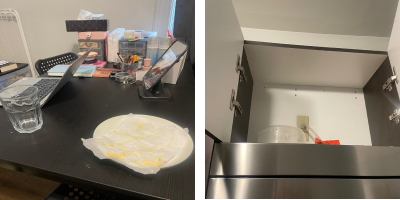}     \end{minipage} & \begin{tabular}[c]{@{}l@{}}If the desk is\\ cluttered\end{tabular} & \begin{tabular}[c]{@{}l@{}}Suggest how to\\ organize items\\ to tidy the desk\end{tabular} \\ \midrule[0.5pt]\vspace{0.4mm}
\begin{tabular}[c]{@{}l@{}}Helps manage the\\ family's nutrition (P1)\end{tabular} & \begin{minipage}{.25\textwidth}       \includegraphics[width=\textwidth]{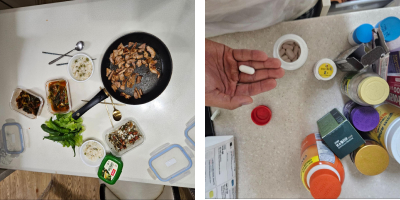}     \end{minipage} & \begin{tabular}[c]{@{}l@{}}After analyzing \\ the food and\\ supplements consumed\\ by family members\end{tabular} & \begin{tabular}[c]{@{}l@{}}Advise on\\ which nutrients are\\ lacking and suggest\\ appropriate supplements\end{tabular} \\ \midrule[0.5pt]\vspace{0.4mm}
\begin{tabular}[c]{@{}l@{}}Recommends outfits\\ based on the weather\\ and harmony (P7)\end{tabular} & \begin{minipage}{.25\textwidth}       \includegraphics[width=\textwidth]{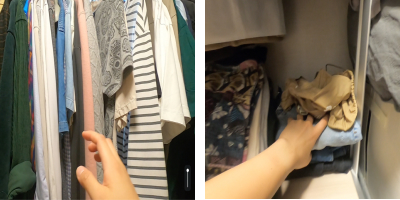}     \end{minipage} & \begin{tabular}[c]{@{}l@{}}If I'm selecting clothes \\ n the dressing room\end{tabular} & \begin{tabular}[c]{@{}l@{}}Recommend the\\ remaining outfit pieces\\ based on the weather\\ and overall harmony\end{tabular} \\ \midrule[0.5pt]\vspace{0.4mm}
\begin{tabular}[c]{@{}l@{}}Analyzes the\\ cooking process\\ and provides tips (P4)\end{tabular} & \begin{minipage}{.25\textwidth}       \includegraphics[width=\textwidth]{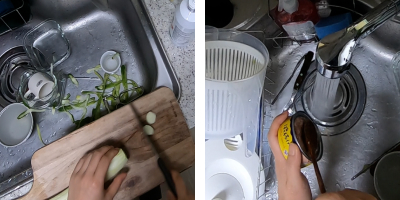}     \end{minipage} & \begin{tabular}[c]{@{}l@{}}If inefficient or unhygienic\\ moments are detected\\ during cooking\end{tabular} & \begin{tabular}[c]{@{}l@{}}Provide advice on\\ improving the\\ cooking process\end{tabular} \\ \midrule[0.5pt]\vspace{0.4mm}
\begin{tabular}[c]{@{}l@{}}Analyzes sleeping posture\\ and offers\\ suitable advice (P12)\end{tabular} & \begin{minipage}{.25\textwidth}       \includegraphics[width=\textwidth]{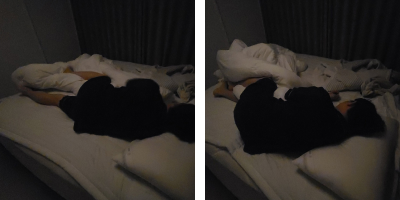}     \end{minipage} & \begin{tabular}[c]{@{}l@{}}If poor sleeping\\ postures are\\ detected during sleep\end{tabular} & \begin{tabular}[c]{@{}l@{}}Upon waking,\\ suggest stretches or\\ foods that benefit the\\ affected body part\end{tabular} \\ \bottomrule[1pt]\vspace{0.4mm}
\end{tabular}
\end{table*}

Participants aimed to go beyond simple convenience by developing features that analyze situations, identify root causes, and propose solutions. These features were intended to detect previously unrecognized problems and provide immediate support for their resolution (Table 4).

They leveraged camera sensors' analytical capabilities to identify unnoticed issues and explore ways to improve their daily lives. For example, P4 developed a feature to analyze their cooking routine, identifying inefficient movements and potential hygiene risks. Similarly, participants created features to analyze specific problem situations, aiming to identify their root causes and facilitate effective solutions.
\begin{quote}\textit{
    I’ve had a few instances where I woke up feeling stiff, so I thought it would be helpful to analyze whether I’m adopting any bad sleeping postures during the night. (P12)} 
\end{quote}

Instead of merely diagnosing situations, participants leveraged the VLM-based camera sensor to develop smart home features that identified problem causes and proposed solutions. For example, P9 developed a feature that suggests tidying up distracting items before starting a study session, while P10 created a feature that provides personalized advice for more effective tidying and organization.

They expressed satisfaction with the advisory features, appreciating their ability to deliver tailored, professional-like guidance specific to their home environment. A key advantage of these features was their capacity to provide personalized, actionable solutions that accounted for the current state of household items, available ingredients, furniture layout, and appliances. 
\begin{quote}\textit{
    I’m not good at organizing, so even though I have enough storage space, I often don’t know where to put things. […] The information from YouTube or blogs doesn’t help much because their spaces are structured differently from mine. However, the feature that my home could be analyzed to give me customized storage advice was the best part. (P10)}
\end{quote}
\begin{quote}\textit{
    I wanted to create features that didn’t just turn lights on and off but could offer fundamental advice, like telling me what was wrong and how I could fix it to make things better. (P4)}
\end{quote}

In conclusion, participants aimed to create smart home features that function as advisors, helping them identify opportunities to improve daily habits and receive personalized advice to enhance their quality of life.

\subsection{How did the characteristics of VLM camera sensors impact the DIY process?}

This section outlines the key characteristics of VLM-based camera sensors and explains the new potential they introduced in the DIY setup process, along with the important factors that need to be considered.

\subsubsection{Comprehensive Sensing: Building Features That Adapt Flexibly to Various Situations}

VLM camera sensors showed the capability to understand the current context without user intervention. Building on this capability, participants constructed more comprehensive rules. The potential of these comprehensive rules led participants to focus less on the specifics of each feature and instead increasingly consider the “fundamental goals of the smart home.”
\begin{quote}\textit{
     I didn’t have to build separate features for each situation, like when my dog howled or chewed on the sofa. The ability to cover many situations under a single feature, such as “Dog Care Alert,” was a major advantage. (P11)}
\end{quote}

This approach enabled a flexible smart home feature that reduced the need for participants to specify every scenario while allowing the system to effectively respond to situations the user may have overlooked.
\begin{quote}\textit{
    I had to specify each risky situation before, but now, using a command like ‘alert me for any dangerous situation’ covers risks I hadn’t even considered. (P1)}
\end{quote}

However, developing smart home features using comprehensive rules presented a challenge. When creating comprehensive features covering multiple situations, it was critical to carefully adjust the rule’s level to ensure it was broad enough to address diverse situations while accurately reflecting user intent.
\begin{quote}\textit{
    When I asked it to analyze my style, I expected it to include my hairstyle, but it only analyzed my fashion. So, I had to adjust it to include both. (P8)}
\end{quote}

In conclusion, VLM camera sensors offer the potential to build smart home features that adapt flexibly to various situations. However, successful implementation required careful consideration of how to set the command level.

\subsubsection{Indirect Sensing Through Inference: Using Unobserved Factors for Feature Execution}

Through inference capabilities, VLM camera sensors held the potential to infer elements that were not directly captured on camera, further expanding their sensing role. Leveraging this capability, participants used elements not captured on the image as sources for executing features. 
\begin{quote}\textit{
    Dust and hair are hard to see, and monitoring the entire floor isn’t realistic either, so instead, I set it up to infer where the floor is dirty based on human activities. For example, if someone dries their hair, it suggests cleaning that area. (P3) - Figure 12}
\end{quote}

\begin{figure}
    \centering
    \includegraphics[width=1\linewidth]{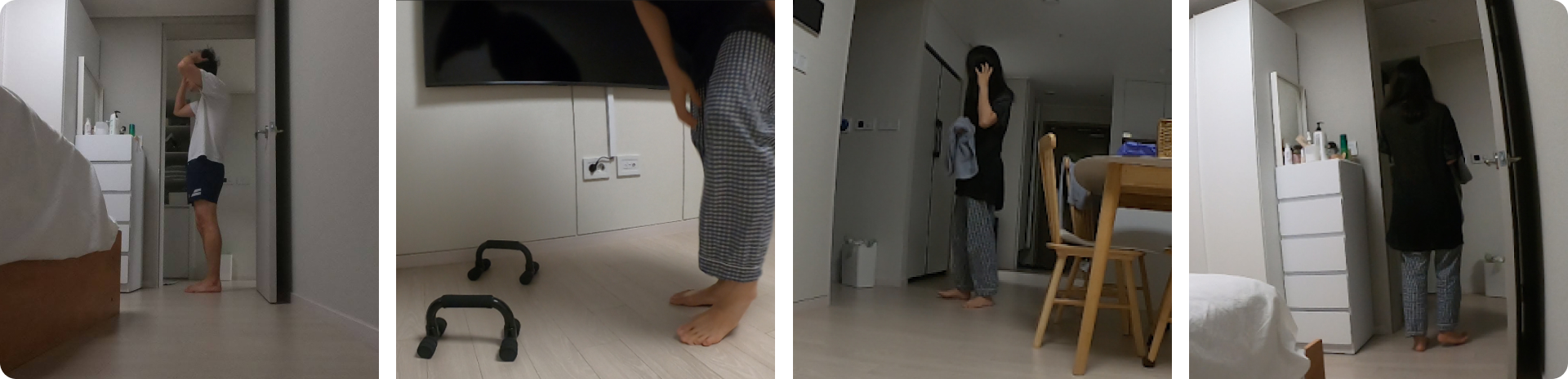}
    \caption{Photos of P3 used to infer a dirty space based on human activities}
    \label{fig:enter-label}
\end{figure}

The ability to infer unseen elements meant that participants did not need to capture every important detail on camera to use smart home features. This potential also provided significant advantages in terms of efficiency and privacy.
\begin{quote}\textit{
    To be honest, it feels burdensome when my facial expressions are being recorded while I’m working. Instead, I thought having the camera analyze my behaviors to understand my emotions would be a better way. (P3) - Figure 13-(a)}
\end{quote}
\begin{quote}\textit{
    For the feature that estimates my calorie expenditure while squatting, using a first-person perspective is more practical than filming from a third-person view. It protects my privacy by not capturing my appearance and allows me to work out anywhere with just one sensor. (P9) - Figure 13-(b)}
\end{quote}

\begin{figure}%
\subfloat[]{{\includegraphics[width=0.47\textwidth ]{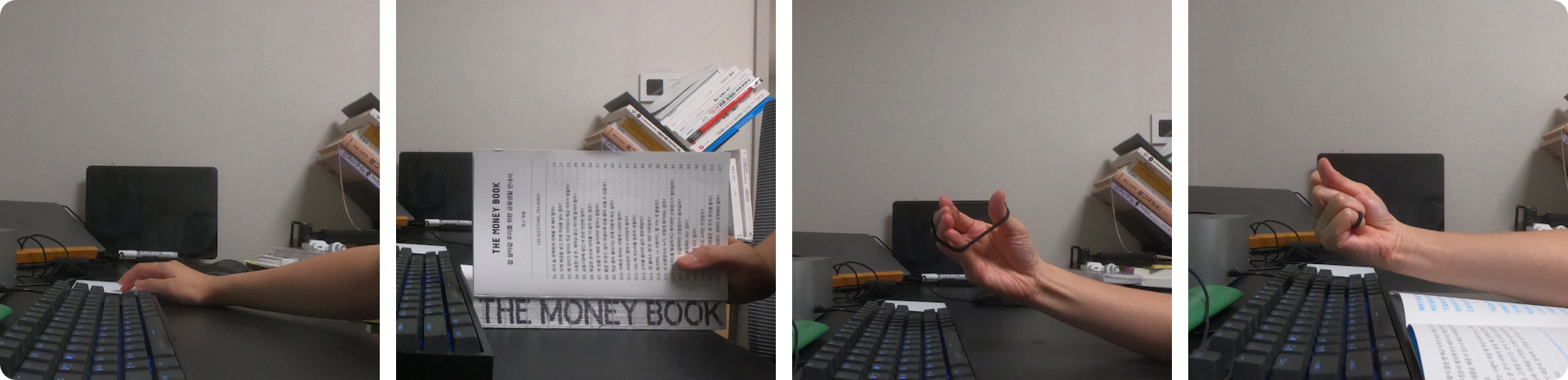}}}%
\hfill
\subfloat[]{{\includegraphics[width=0.47\textwidth ]{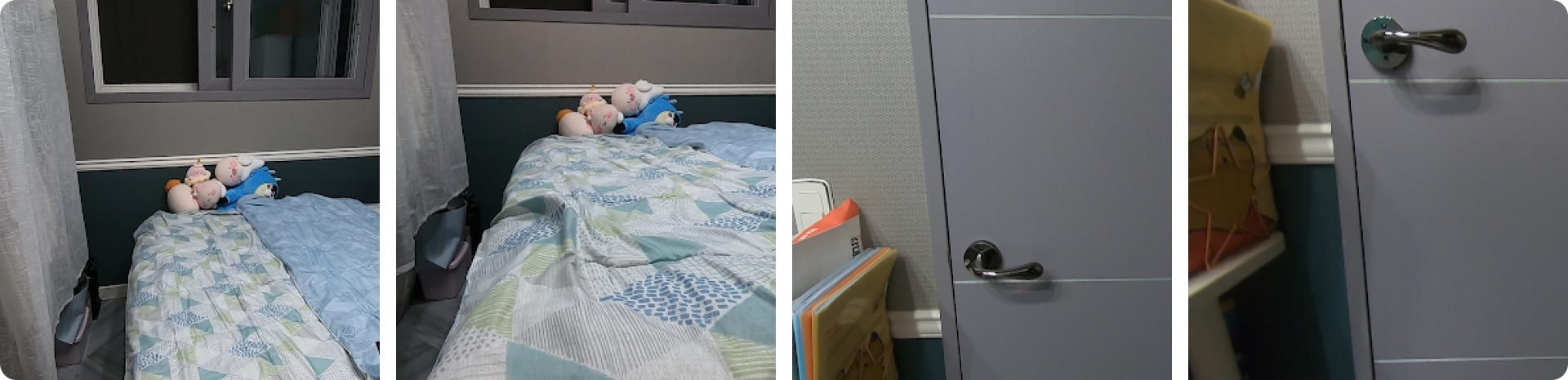}}}%
\caption{Participants' photos, (a) P3 developed a feature to infer emotions such as focus, nervousness, or anger based on hand movements like opening a book, playing with a rubber band, or tightly clenching her fist, (b) P9 developed a feature to count squats and calculate calories burned by tracking changes in her field of view}%
\end{figure}

However, due to the sensor's inference capabilities, participants experienced confusion about what the sensor could and could not detect. At times, the sensor even provided fabricated or false information, which led to a decline in participants' trust.
\begin{quote}\textit{
    One time, the system said there was food waste in the sink when the sink wasn’t even visible in my images. That made me question whether all the meaningful inference and helpful guidance it had given me before was a lie. I felt really betrayed. (P4)}
\end{quote}

In conclusion, the VLM camera sensor demonstrated the potential to expand its role by inferring unseen elements. While this capability offered significant benefits in terms of efficiency and privacy, it also blurred the lines of what the sensor could reliably sense, making it harder for participants to differentiate between its capabilities and limitations.

\subsubsection{Perspective-embodied Sensing: Enabling Situation Analysis from Diverse Perspectives}

Participants were able to utilize various types of camera sensors when building a single feature, enabling them to sense a situation from user-focused, device-focused, and spatial-focused perspectives depending on the configuration of the camera sensor they selected. The VLM camera sensor, taking into account these diverse perspectives, was able to analyze situations in a way that reflected the chosen perspective. In the user-focused perspective (e.g., wearable cameras), the sensor recognized that the scene captured on the screen represented the participant’s current view, allowing it to focus on the objects or elements that the participant was observing or interacting with. In the device-focused perspective (e.g., cameras embedded in devices), the context of the device's purpose and function was considered in the analysis. For example, from a refrigerator's perspective, the camera's role could be focused on monitoring food freshness, tracking inventory, and supporting automated grocery shopping or meal planning. For the spatial-focused perspective (e.g., 360-degree cameras), the analysis went beyond the user-centered approach seen in wearable cameras, taking into account the broader contextual surroundings, such as the overall atmosphere of the home. This diversity in perspective enabled the sensor to interpret the home situation in a more contextually rich manner. 
\begin{quote}\textit{
    Since the camera attached to the mirror doesn’t capture the mirror itself, it just looks like a person is standing still in the image. So, I had to specify that this image was taken by a mirror-embedded camera, which helped the sensor better understand that the person standing in the center was actually checking their outfit in the mirror. (P8)} 
\end{quote}

Furthermore, the ability to explicitly define the camera sensor's perspective allowed participants to assign both physical perspectives (user, device, or spatial) and conceptual perspectives, such as those of a mother, an expert, or a pet. 
\begin{quote}\textit{
    We only see as much as we know. Since I’m not a plant expert, I thought that analyzing it using a plant expert's analytical approach could help identify elements I might have missed and support the diagnosis of the plant’s condition. (P5)} 
\end{quote}

On the other hand, relying on a single camera perspective was sometimes insufficient to capture all critical information. To address this, participants recognized the need for multi-perspective sensing and envisioned collaboration between multiple cameras. However, developing such a system required clear rules and conditions to determine when and how camera perspectives should switch. Participants expressed the need for support in defining these rules to ensure smooth and effective camera transitions. 
\begin{quote}\textit{
    To understand what’s happening with the baby, the camera needs a wide-angle view, but if the baby’s facial expressions are important, it’ll need a close-up. It’s a dilemma. So, I thought about using both cameras together—a wide-angle camera to capture the overall situation and a wearable camera to focus on the details. [...] But in that case, I would need to set rules for when to switch between the cameras, and that could make things a bit more complicated. (P8)}
\end{quote}

In conclusion, VLM camera sensors enabled participants to analyze a single situation from multiple perspectives. Notably, when a single feature required the collaboration of cameras with different perspectives, participants emphasized the need for support in defining rules and conditions for seamless camera switching. 

\subsubsection{Unbounded Sensor Values: Enjoying Unexpected Features}

Unlike commonly used IoT sensors, which produce results within a limited range of values, VLM sensors offer the possibility of generating an unlimited range of outputs. Since everyday situations can vary greatly, the results generated by the camera sensor extend far beyond a fixed scope. Initially, participants tried to control this infinite potential by restricting the range of sensor values to predefined lists. However, this approach often ended up overly limiting the sensor’s capabilities.
\begin{quote}\textit{
    I tried limiting the responses to just two options: either the child is trying to climb the stairs or not. [...] When it came to analyzing what the kids were doing in the living room, I wanted to leave the responses open, as there could be a wide variety of situations. (P6)}
\end{quote}

One of the key benefits of having unrestricted sensor values was the possibility of features being executed in ways even the participants who implemented the feature had not anticipated. In commonly used IoT-based smart homes, a feature could not operate beyond the situations that they directly set, making this an entirely new experience that VLM camera sensors could offer.
\begin{quote}\textit{
    In smart homes, I’m usually less impressed since I create all the features. With IoT, it’s more like, ‘oh, it works as expected.’ But with cameras, the system can execute features I didn’t anticipate, like identifying inefficient habits I wasn’t even aware of and offering helpful advice—that’s a big advantage. (P12)}
\end{quote}

However, this characteristic also introduced new challenges in the process of building smart home features. Since sensor values were not limited to a few predefined lists, it became difficult to follow the traditional trigger-action paradigm. Instead, participants had to set the sensor values as variables and configure actions to execute automatically based on those values. For example, P3 requested that the system automatically select and play suitable music based on her detected emotional state. Also, as the range of possible actions increased, participants grew concerned about features they had not anticipated. To address these concerns, some participants wanted the smart home system to propose potential scenarios where the features could be applied.

In conclusion, VLM camera sensors opened up the possibility of executing features that participants had not anticipated. However, this required building features based on variables, and participants needed additional support to help them account for the wide range of possible scenarios where features could be triggered.

\subsubsection{Interpretation Beyond Sensing: Building Features Without Explicit Definitions}

Participants expected that VLM camera sensors could go beyond merely recognizing current situations and perform additional interpretative tasks. They built smart home features using the VLM camera sensor’s ability to (1) provide useful information, (2) make judgments about the current state, (3) draw comprehensive conclusions, and (4) offer recommendations. As described in Section 4.1, this capability contributed to building diverse smart home features.

Participants perceived the VLM camera sensor as more than just a tool for sensing their home—it was capable of "thinking" based on its observations. By moving beyond simple sensing to interpreting situations, these sensors enabled participants to build complex features without explicit definitions, streamlining processes and allowing for capabilities that were previously impossible.
\begin{quote}\textit{
    If I were to build a feature with an IoT sensor to notify me when my desk is dirty, I’d be stuck trying to define what ‘dirty’ means. But with the camera sensor, it can look at the state of my desk and judge whether it’s dirty or not. This made building the feature much simpler. (P10)}
\end{quote}

However, the sensor’s ability to interpret situations independently sometimes resulted in differences between human and camera sensor interpretations. As a result, participants had to refine and align these differences. To address this, they employed strategies that involved sharing personal information with the sensor, helping it better understand them.
\begin{quote}\textit{
    While the posture warning feature alerts me to things like resting my chin on my hand or shaking my leg, those don’t seem critical to me. However, a severely slouched posture is important because I have back problems, so I made sure to inform the system about my bad back. (P4)}
\end{quote}
\begin{quote}\textit{
    My air conditioner gives off a musty smell, so we have to briefly open and close the door when we first turn it on. But GPT wouldn’t understand that and would probably flag it as wasted energy, so I had to let it know to exclude that situation. (P1)}
\end{quote}

In conclusion, while VLM camera sensors contributed to building a variety of smart home features by interpreting context beyond simple recognition, participants needed support to align sensor interpretations with human expectations during development.

\subsection{What Are the Concerns Regarding VLM Camera Sensor-Based DIY Smart Homes?}

Through this study, we were able to gather users’ concerns as well. Since using a camera sensor requires installing cameras inside the home, participants expressed concerns related to privacy issues (Section 4.3.1). Also, the video footage could not only be recorded but also analyzed, raising new concerns that surpassed the issues traditionally associated with basic home cameras (Sections 4.3.1–4.3.4). In interviews, 5 out of 12 participants (P1, P6, P7, P10, P12) explained that beyond simply recording footage, the analysis of this footage led to additional concerns. This section will discuss four key concerns raised by participants.

\subsubsection{Concerns about Privacy Issues from Full Digitalization of Home Life}

Over the three-week period, participants raised various privacy concerns, many of which aligned with those identified in previous studies on home cameras \cite{jin2022exploring, chalhoub2021did, jin2022exploring}. Specifically, participants emphasized the need for design solutions that go beyond ensuring technical security, highlighting the importance of psychological comfort. Accordingly, they stressed the need for features that clearly indicate when the camera is operating and provide easy control over its functions. 

Moreover, VLM camera sensors introduced an additional layer of concern, as they not only capture video but also analyze it. Participants were particularly worried that the analysis and digitization of images could expose sensitive information, raising privacy issues. They feared that unintended details might be detected, analyzed, and used as materials for smart home features, potentially leading to the overexposure of their private lives.
\begin{quote}\textit{
    When I take photos on my phone, they automatically update on my Nest Hub. Seeing that those photos are analyzed made me think, ‘this is a bit of an invasion of privacy.’ If the system can reference those photos, it could figure out where I’ve been or what I did outdoors, which felt a bit uncomfortable. (P1)}
\end{quote}
\begin{quote}\textit{
    If I’m having a private conversation with a friend, and I find out that the messages are being continuously analyzed, it wouldn’t feel great. (P10)}
\end{quote}

Participants also raised concerns about the potential commercial use of information collected inside their homes. They feared that their personal data could be misused to generate indiscriminate product recommendations benefiting corporations, which heightened their unease about privacy and data exploitation.
\begin{quote}\textit{
    It could recommend clothes or detergent I need, but if I think about it negatively, they could be using my data for advertising, which would end up disrupting my daily life. (P7)}
\end{quote}

In summary, participants expressed concerns about the digitalization of their daily lives through camera sensors, particularly regarding the unintended analysis of personal information and its use in smart home features, as well as the potential commercial exploitation of their data. These concerns highlighted the need for transparency and user control over how their information is processed and utilized. 

\subsubsection{Concerns About Replacing Interactions Between Family Members}

Some of the features participants built had the potential to replace existing interactions between family members. P2 explained that he considered introducing features to avoid nagging their child, while P1 wanted to help their parents take their supplements correctly.
\begin{quote}\textit{
    My daughter often reads or draws without turning the light on. When I noticed later, I ended up nagging her, so I thought it would be helpful if the light turned on automatically in real time. (P2)}
\end{quote}
\begin{quote}\textit{
    My parents are taking all the supplements they have without much thought. So, I thought it would be helpful if the smart home could track what they’ve eaten today and provide information on which supplements they still need to take. (P1)}
\end{quote}

However, participants also expressed concern that such a feature could reduce social interactions within the family.
\begin{quote}\textit{
    When my sister forgets her car key and leaves the house, she usually calls me, and I have to look for it, which leads to a bit of bickering. But if everything worked perfectly, even those small interactions could disappear, and that’s something I was concerned about. (P8)}
\end{quote}
\begin{quote}\textit{
    Features like guiding my parents to take their supplements correctly or informing them if something shouldn’t go in the fridge could be useful when I’m not home. But when I’m there, I think they’d rather hear it from me than from the system. (P1)}
\end{quote}

In summary, participants were concerned that smart home features might reduce meaningful family interactions and wanted to ensure that systems were not designed to diminish these important connections.

\subsubsection{Concerns About Over Dependence on Automated Decision-Making}

Participants explained that even simple features quickly became indispensable once integrated into their routine. They expressed concern that, as more intelligent features are added, they might become overly reliant on them.
\begin{quote}\textit{
    My parents can’t live without the counter sensor now. It’s just a simple feature that automatically turns on the bathroom light, but once we get used to it, we can’t live without it. I think as more smart feature becomes a part of daily life, we’ll end up depending on it even more. (P1)}
\end{quote}

Some participants also worried that smart home features could interfere with their positive habits as smart homes become capable of making automated decisions. 
\begin{quote}\textit{
    We all have our own cleaning routines, but if we become too dependent on the smart home, I might only clean the areas the smart home tells me are dirty. In a way, this could end up disrupting the good habits I’ve already developed. (P1)}
\end{quote}

To address these concerns, some participants suggested that a good solution would be to set the smart home features to activate only when they specifically wanted them.
\begin{quote}\textit{
    I like to think things through as much as possible and try to figure out the answer myself before seeing what others think. So, I’d prefer if the system only gave me answers when I physically pressed a button. (P8)}
\end{quote}

In conclusion, participants wanted to carefully consider when and how these features should operate, ensuring they avoid over-reliance on technology that could negatively affect their well-being.

\subsubsection{Concerns and Discomfort About AI Controlling Daily Life}

All participants mentioned being surprised by the detailed descriptions the AI provided when analyzing images. Some expressed concerns about AI becoming too intrusive in their daily lives.
\begin{quote}\textit{
    We sometimes humanize AI, and the thought of AI watching and analyzing my home—whether on-device or not—feels no different. It’s like a stranger observing my daily life. (P6)}
\end{quote}

Participants particularly felt discomfort when they perceived AI as trying to analyze and excessively improve or control their behavior. This led to stronger feelings of rejection toward the technology.
\begin{quote}\textit{
    AI judging my actions can make me feel like I’m being watched, which creates a sense of discomfort. That’s one of the things I was concerned about. (P12)}
\end{quote}
\begin{quote}\textit{
    When AI analyzed my behavior and suggested that I seemed anxious based on how I was holding a can, it felt like it was constantly trying to judge and correct me, which made me really uncomfortable. (P3)}
\end{quote}

Participants were particularly wary of how AI control could steer their daily lives in unintended directions or negatively influence their decisions. 
\begin{quote}\textit{
    If it provides information tailored to my situation, I feel like I have to follow that advice, even though I wouldn’t know if it’s correct. If it were to suggest actions that are wrong or harmful to me, it could become a seriously threatening experience for my life. (P10)} 
\end{quote}

In summary, participants expressed concerns about AI becoming too intrusive in their lives, especially when it appeared to judge or correct them. This sense of being monitored and controlled by AI caused discomfort and unease.

\section{Discussion \& Design Implications}

\subsection{DIY Smart Home Construction Using VLM Camera Sensors}

According to our study, the process of building DIY smart homes using VLM camera sensors required new considerations and additional support (Section 4.2). These sensors can understand home situations without users manually interpreting sensor data, simplifying parts of the DIY process. However, this also raised the possibility of the system behaving unexpectedly, even with desired features in place. This section summarizes key considerations for maintaining user control when building DIY smart home features with VLM sensors. Based on the five stages of DIY smart home construction \cite{yun2023potential}, we propose design implications for the support users need at each stage when using VLM sensors.

\subsubsection{Step 1. Selecting Sensing Situations and a Sensor: Proposing Higher-Level Feature}

As mentioned in Section 4.2.1, VLM camera sensors allow more comprehensive commands, like "Set lighting for desk activities," simplifying setup, and enabling smart homes to flexibly adapt to various household situations. However, overly broad goals can hinder smooth communication with camera sensors, making it essential to help users set clear goals and define the scope of situations to be included in those goals (Section 4.2.1).

To assist users, the smart home system can suggest higher-level features when the user inputs the feature they want to build. For example, if a user wishes to create a feature like “plays an ‘It’s okay’ message when the dog howls,” the system might propose, “Shall we set up an environment to soothe your dog when she feels anxious?” At the same time, to ensure the user retains control over the system, it is important to explain the various situations that may be included when selecting a higher-level feature. This allows the user to verify that all desired contexts are covered and make adjustments to the definitions if necessary. In this way, a smart home system using VLM camera sensors should help users explore appropriate levels of features.

\subsubsection{Step 2\&3. Defining Role and Location of Sensor: Framing the Focus for Scene Analysis}

Commonly used IoT sensors provide objective and consistent results by delivering fixed numerical values under identical conditions. In contrast, VLM-based camera sensors can analyze the same scene from multiple perspectives, depending on how users define the sensor's role. As noted in Section 4.2.3, scene interpretation varies based on the selected perspective. Participants also explored conceptual perspectives (Section 4.2.3) and interpreted situations based on elements that indirectly reveal the context of a situation (Section 4.2.2), enabling deeper and more meaningful analysis of home situations. However, this flexibility to freely define and change the sensor's role also revealed certain limitations. Participants found it difficult to distinguish between what the sensor could and could not detect (Section 4.2.2). Additionally, they required support for smooth and seamless perspective switching when multiple perspectives were needed (Section 4.2.3). 

To support the process of defining the sensor's role, it may be beneficial to present users with various perspectives through which the sensor can analyze home situations. Visualizing how the sensor interprets situations and allowing users to modify this logic could further enhance their control and understanding. Additionally, if a switch between the presented perspectives is required, users should be provided with tools to easily define and adjust the logic for perspective switching. This support would help users set realistic expectations and facilitate a more meaningful analysis of home situations. 

\subsubsection{Step 4. Creating Rules for Features: Supporting the Logic-Setting Process for Autonomous Action Selection}

In building commonly used IoT sensor-based features, the Trigger-Action paradigm allowed users to create specific rules about which actions to execute when certain sensor conditions were met \cite{brackenbury2019users, ur2014practical, zhao2021understanding}. However, as noted in Section 4.2.4, VLM camera sensors can output a wide range of sensor values. This characteristic requires a new rule-setting approach where sensor values are treated as variables, and actions are determined logically based on them. For example, a feature that “adjusts the environment based on desk activities” could trigger different actions depending on whether the goal is focus, energy conservation, or fatigue prevention. Therefore, it is necessary to set rules that determine which automated actions to execute based on the users’ customized logic.

To assist in this process, instead of assigning actions for each sensor value, the smart home system should autonomously decide actions based on sensor data, while helping users define the underlying logic. For example, if the user sets a feature to 'adjust the environment based on desk activities,' the system could prompt the user to specify the logic behind adjusting the environment. Additionally, to avoid confusion when unexpected features occur (Section 4.2.4), the system could provide scenario examples, allowing users to adjust them as needed. Through this approach, smart home systems should ensure autonomy in a way that aligns with the user’s intentions.

\subsubsection{Step 5. Testing and Modifying Features: Supporting Meaningful Personal Information Sharing Interactions}

When building features with commonly used IoT sensors, the system follows simple, user-defined rules, so unexpected results usually indicate a rule error. In contrast, with VLM camera sensors, even correctly set rules can lead to discrepancies between expected and actual outcomes (Section 4.2.5). These differences occur due to AI interpretation, requiring adjustments when the AI's conclusions differ from the user's expectations.

To bridge the gap between AI’s interpretation and the user’s perception, it is crucial to provide ways to reconcile these differences without overwhelming the user. One approach is to design interactions that encourage users to reflect on their daily lives, creating meaningful experiences while providing better information for the smart home system. As mentioned in Section 4.1, some participants saw the camera sensor’s analysis as an opportunity to gain an objective view of their home and behavior. Comparing this objective view with their own perspective can foster valuable self-reflection and also give the smart home system a chance to learn more about the user. In conclusion, effectively using VLM camera sensors may require additional information to align the system with the user’s perspective. Smart home systems should support interactions that facilitate meaningful sharing of personal information.

\subsection{Living with Intelligence}

Many previous studies have explored the core values that users expect from smart homes. Although terminology may vary slightly, these studies consistently highlight three key values. First, the value of “\textbf{protection},” which involves ensuring home security and monitoring for intruders. Second, the value of “\textbf{productivity},” is achieved through the automation and remote control of home appliances. Third, the value of “\textbf{pleasure},” gained from creating environments tailored to household activities \cite{strengers2019protection, jensen2018designing, wozniak2023inhabiting}. Building on these studies, this section discusses how the three smart home roles identified in this study (Section 4.1) may expand existing values.

\subsubsection{Beyond protection to an auto-monitoring role: Monitoring the home from various perspectives}

Ensuring home safety, monitoring intruders, and remotely observing and caring for vulnerable individuals have long been key concerns for smart home users. While traditional home cameras automated certain monitoring tasks, such as detecting the approach of strangers, package deliveries, and pet barking, they still required users to manually monitor video feeds. In contrast, the introduction of VLM camera sensors enabled users to develop features that automatically recognize home situations and deliver relevant information without the need for direct monitoring (Section 4.1.1). Depending on the household's characteristics and needs, the targets and scenarios for automatic monitoring varied, but VLM camera sensors provided the flexibility to automatically monitor specific situations that users considered important. This shift allowed participants to move beyond basic surveillance of external threats and develop features that recognize and analyze a broader range of home situations. For example, participants can build features that quantify the state of home organization, identify significant changes in the home compared to a year prior, and reveal habitual behavior patterns that users had not previously recognized. By supporting these features, VLM camera sensor-based smart homes extend their role beyond mere protection to facilitating a deeper understanding of users' daily routines and behaviors. This shift allows users to gain new insights into their daily lives, fostering opportunities for personal growth and lifestyle improvement. By encouraging self-reflection and promoting a better understanding of personal habits and routines, VLM camera sensors could contribute to enhancing the quality of everyday experiences. 

\subsubsection{Beyond productivity to an assistant role: Enriching daily experiences}

Productivity is one of the core values of smart homes, enabling them to save time and improve efficiency by automatically operating appliances at optimal times. However, previous studies have expressed concerns that prioritizing efficiency alone could obscure other important household values \cite{berger2023accidentally, alves2021collection, ehrenberg2021technology, strengers2022isn}. In response, assistant role-based smart home features (Section 4.1.2) offer possibilities beyond productivity by providing richer support for user activities. VLM camera sensor-based features provide support by detecting subtle changes or by providing useful, context-relevant information. For example, they can help users detect subtle changes in plants, enhancing their sense of accomplishment in plant care, or analyze the non-verbal behaviors of babies and pets to help users better understand and respond to their needs. Assistant features can also make undesirable tasks more engaging by incorporating playful or motivational elements. During cooking, the system could provide interesting facts or the origin of ingredients to make the activity more enjoyable. While cleaning, it could offer motivational feedback like, "You’ve cleaned 20\% more area today compared to yesterday," fostering a sense of achievement. These possibilities signal a shift from perceiving smart homes as mere automation tools to recognizing them as systems that enrich daily life, highlighting their potential to provide more meaningful experiences for users.

\subsubsection{Beyond pleasure to an advisory role: Improving the overall quality of life}

The concept of pleasure in smart home features stems from features that create an optimal atmosphere in response to user needs. For instance, features like "movie mode" for film viewing or "party mode" for social gatherings transform the home into a special space, providing emotional satisfaction. However, with VLM camera sensors, participants showed a preference for features that go beyond entertainment, seeking advisory support aligned with deeper personal goals (Section 4.1.3). Recent studies on AI applications in nutrition management, medication adherence, and medical diagnosis \cite{szymanski2024integrating, harrington2022s, li2024conversational} suggest the potential for smart home features to offer professional guidance within the home. This shift reflects users' perception of AI as a tool for supporting well-being rather than merely providing entertainment. Nonetheless, pleasure remains essential for fostering user affinity with technology and maintaining engagement. For example, using playful or humorous notifications instead of directive messages can create a more positive user experience. Since the home is a space for comfort and relaxation, interactive strategies that deliver lifestyle advice in an enjoyable way may help sustain user engagement. Ultimately, designing features that blend advisory support with pleasurable experiences may not only sustain user engagement but also promote users' long-term well-being. This approach highlights the potential of smart home technologies to provide both emotional satisfaction and meaningful personal growth. 

\subsection{Challenges of Living with Intelligence}

As discussed in Section 5.2, VLM camera sensor-based smart homes offer users new value. However, as highlighted in Section 4.3, these advancements may also introduce new concerns. This section examines how the introduction of VLM camera sensors could intensify existing issues in smart homes. 

\subsubsection{Expansion of Privacy Issues}

Privacy has long been a critical issue in the smart home domain, but the introduction of VLM camera sensors introduces new dimensions of privacy threats. Unlike commonly used IoT sensors, VLM camera sensors can extract a wide range of personal information from a single image \cite{venugopalan2024aragorn, xu2024dipa2}, including behavior, facial expressions, lifestyle habits, food preferences, and favorite products (Section 4.3.1). This capability makes it difficult for users to know exactly what information is being collected, increasing the risk of unintentional exposure and weakening their control over personal data \cite{venugopalan2024aragorn, xu2024dipa2}. Moreover, since VLM camera sensors enable a richer analysis of home environments, the misuse of this data poses heightened risks. For instance, users' preference data could be exploited for targeted marketing. More concerningly, personal lifestyle data could be used to influence insurance premiums or hiring decisions \cite{lee2021ai}, leading to tangible societal risks. 

To address these threats, technical safeguards such as blurring irrelevant parts of images \cite{venugopalan2024aragorn} or automatically detecting sensitive information \cite{xu2024dipa2} are being actively explored. However, technical measures alone may not be sufficient to ensure user trust. When a video related to a recent conversation with friends appears as a YouTube recommendation, people often joke, "Was Siri listening to us?"—a remark that reflects underlying suspicion and unease. If this shifts to "Was Siri watching us?", the sense of discomfort is likely to be even greater. Such situations illustrate how users fill in the gaps of technological opacity with their imagination, leading to sustained anxiety regardless of the system's actual behavior. While the simplest solution would be to physically block the camera's view \cite{shalawadi2024manual, cheng2019peekaboo}, this is not a viable option for VLM camera sensors since the camera itself functions as a sensor. A better approach would be to increase transparency regarding what data is being collected and how it is being used. However, processing data in a way that provides users with psychological comfort may not be an easy task. Future research should focus on interaction design and data management strategies that ensure users' psychological comfort and trust. To prevent users' privacy from being indiscriminately sacrificed in favor of smart home benefits, further research and innovative approaches are essential. 

\subsubsection{Replacement of Positive Practices}

Previous studies have consistently highlighted the potential negative impact of smart home automation on user well-being. For example, automated lighting control could reduce physical activity, while task reminders might hinder memory development \cite{strengers2022isn}, and automation in shared spaces could weaken daily practices of mutual care \cite{lee2024we}. Unlike commonly used IoT sensors, VLM camera sensors can recognize and interpret situations (Section 4.2.5), enabling more precise and advanced levels of automation. With this capability, smart homes can now determine when cleaning is needed, monitor if a child requires parental support, or detect inefficient cooking habits. While these features offer greater convenience and efficiency, automated intervention may reduce family interaction (Section 4.3.2) and encourage over-reliance on the system's decisions, weakening users' self-judgment (Section 4.3.3). As a result, essential daily practices that users previously performed independently risk being indiscriminately replaced. 

To address these risks, it is important to inform users about the potential benefits, drawbacks, and ethical implications of the features they create. For instance, when building a feature that detects a baby’s discomfort based on facial expressions and movements, the system could display a message like, “This feature may reduce emotional bonding moments between parents and their child.” Such messages help users recognize the broader impact of automation, enabling them to make more informed choices. Moreover, leveraging VLM camera sensors’ ability to suggest contextually appropriate actions (Section 4.2.4) can support users in making more balanced decisions by offering multiple action options that reflect diverse values beyond efficiency. Our suggestion is also aligned with \cite{challenge} where the author claimed the importance of viewing and designing AI as a provocateur by providing questions and different perspectives to enable users' critical thinking instead of automating and assisting situations for efficiency. We think that the VLM-enabled smart home context is one of the urgent domains where this type of approach should be seriously explored. This approach allows users to design a smart home experience that balances not only convenience and efficiency but also values like social interaction, self-determination, and well-being. 

\subsubsection{Distorted Power of AI in the Home}

To address the issue of smart home systems executing unexpected automated actions that reduce user control and trust, previous studies have explored strategies to enhance the explainability of smart home features \cite{dai2023effect, das2023explainable}. These efforts aim to make the decision-making process transparent, enabling users to better understand system behavior. However, the introduction of VLM camera sensors adds a new layer of complexity beyond simple explainability. If issues like bias, discrimination, hallucination, and filter bubbles — which have been widely discussed in the context of online AI systems — are reflected in smart home systems, they could directly influence users' thoughts and behaviors in undesirable ways (Section 4.3.4). For example, AI bias might unfairly restrict certain actions, filter bubbles could reinforce existing preferences, and hallucinations might cause users to accept false information as true. This poses the risk of AI directly controlling users’ behavior and decision-making (Section 4.3.4), raising the possibility that AI could exert distorted power within the home. 

To address these concerns, smart home systems must be designed to minimize AI's influence on users' daily lives and ensure users maintain control over their own decisions. Even when AI systems observe daily activities and offer recommendations, users should have the ability to exercise independent judgment, with smart home systems promoting critical thinking and personal choice rather than imposing specific behaviors. This approach is crucial to prevent the unfair exercise of AI power and to create a truly "smart" home environment that supports users' autonomy and well-being. These considerations underscore the need for future research to define the appropriate scope of AI's influence, ensuring that advisor features (Section 4.1.3) genuinely enhance users' lives.

\section{Limitations \& Future Work}

This study employed a diary study-based experience prototyping method, where participants simulated a DIY smart home setup to minimize the indiscriminate exposure of home image data. However, real-world deployment of camera sensors may introduce additional challenges beyond those identified here. For example, while participants manually selected key images during the simulation, real-world implementations would require this process to be fully automated. Additionally, determining which images are suitable for analysis may involve other complexities that were not considered in this study. Moreover, a fully implemented and integrated intelligent smart home system could give rise to new and unforeseen issues. For example, frequent system errors might shift users' expectations and reduce their trust in the smart home’s role and reliability. Future research should address these challenges by employing on-device models capable of real-time analysis. This approach would offer a more accurate understanding of the practical issues that arise during real-world usage, providing deeper insights into user experience.

Moreover, this study involved one representative participant per household for the DIY smart home setup, focusing on those most interested in smart homes and responsible for maintaining and updating smart home features. However, building a shared smart home requires addressing the diverse needs of all household members \cite{koshy2021we, garg2022social, lee2024we}. Based on the findings and design implications of this study, future research could further develop a toolkit for DIY smart home setups with VLM camera sensors, emphasizing the collaborative process of multiple users building a smart home together \cite{xue2024should}.

\section{Conclusion}

This study explored new possibilities and considerations in DIY smart home building as VLM technology advances, enabling camera sensors to autonomously understand home situations. We conducted a three-week diary-based experience prototyping with 12 participants, guiding them to simulate DIY smart home-building processes using VLM camera sensors. According to our findings, participants built features for auto-monitoring, assistant, and advisory roles. Unlike commonly used IoT sensors, VLM camera sensors introduced new characteristics—comprehensive sensing, indirect sensing through inference, perspective-embodied sensing, unbounded sensor values, and interpretation beyond sensing—which brought new possibilities and considerations to the DIY process. Participants also expressed concerns related to home cameras and their use as sensors. Based on these findings, we proposed design implications for constructing DIY smart homes with VLM camera sensors and discussed how VLM camera sensor-based smart homes may expand existing values while exacerbating current challenges. Through this study, we hope to contribute to future research that explores meaningful home intelligence while prioritizing user autonomy and control.

\begin{acks}
We thank all participants in our study. We also thank the anonymous reviewers for their insightful comments and suggestions regarding this paper. This work was supported by the National Research Foundation of Korea(NRF) grant funded by the Korea government(MSIT) (No. NRF-2021R1A2C2004263).
\end{acks}

\appendix

\bibliographystyle{ACM-Reference-Format}
\bibliography{sample-base}


\begin{thebibliography}{73}


\ifx \showCODEN    \undefined \def \showCODEN     #1{\unskip}     \fi
\ifx \showDOI      \undefined \def \showDOI       #1{#1}\fi
\ifx \showISBNx    \undefined \def \showISBNx     #1{\unskip}     \fi
\ifx \showISBNxiii \undefined \def \showISBNxiii  #1{\unskip}     \fi
\ifx \showISSN     \undefined \def \showISSN      #1{\unskip}     \fi
\ifx \showLCCN     \undefined \def \showLCCN      #1{\unskip}     \fi
\ifx \shownote     \undefined \def \shownote      #1{#1}          \fi
\ifx \showarticletitle \undefined \def \showarticletitle #1{#1}   \fi
\ifx \showURL      \undefined \def \showURL       {\relax}        \fi
\providecommand\bibfield[2]{#2}
\providecommand\bibinfo[2]{#2}
\providecommand\natexlab[1]{#1}
\providecommand\showeprint[2][]{arXiv:#2}

\bibitem[Abdallah et~al\mbox{.}(2017)]%
        {abdallah2017lessons}
\bibfield{author}{\bibinfo{person}{Raef Abdallah}, \bibinfo{person}{Lanyu Xu}, {and} \bibinfo{person}{Weisong Shi}.} \bibinfo{year}{2017}\natexlab{}.
\newblock \showarticletitle{Lessons and experiences of a DIY smart home}. In \bibinfo{booktitle}{\emph{Proceedings of the Workshop on Smart Internet of Things}}. \bibinfo{pages}{1--6}.
\newblock


\bibitem[Aiswarya et~al\mbox{.}(2021)]%
        {aiswarya2021living}
\bibfield{author}{\bibinfo{person}{R Aiswarya}, \bibinfo{person}{Vaishali Anil}, \bibinfo{person}{Jayasree Raveendran}, \bibinfo{person}{Gijs Van~Wijk}, {and} \bibinfo{person}{Shantanu Talukdar}.} \bibinfo{year}{2021}\natexlab{}.
\newblock \showarticletitle{Living as a Service (LaaS): An Experiment to Evaluate Living Experience in Connected Homes in the Netherlands.}. In \bibinfo{booktitle}{\emph{CHI Extended Abstracts}}. \bibinfo{pages}{360--1}.
\newblock


\bibitem[Alves-Oliveira et~al\mbox{.}(2021)]%
        {alves2021collection}
\bibfield{author}{\bibinfo{person}{Patr{\'\i}cia Alves-Oliveira}, \bibinfo{person}{Maria~Luce Lupetti}, \bibinfo{person}{Michal Luria}, \bibinfo{person}{Diana L{\"o}ffler}, \bibinfo{person}{Mafalda Gamboa}, \bibinfo{person}{Lea Albaugh}, \bibinfo{person}{Waki Kamino}, \bibinfo{person}{Anastasia K.~Ostrowski}, \bibinfo{person}{David Puljiz}, \bibinfo{person}{Pedro Reynolds-Cu{\'e}llar}, {et~al\mbox{.}}} \bibinfo{year}{2021}\natexlab{}.
\newblock \showarticletitle{Collection of metaphors for human-robot interaction}. In \bibinfo{booktitle}{\emph{Proceedings of the 2021 ACM Designing Interactive Systems Conference}}. \bibinfo{pages}{1366--1379}.
\newblock


\bibitem[Berger et~al\mbox{.}(2023)]%
        {berger2023accidentally}
\bibfield{author}{\bibinfo{person}{Arne Berger}, \bibinfo{person}{Albrecht Kurze}, \bibinfo{person}{Andreas Bischof}, \bibinfo{person}{Jesse~Josua Benjamin}, \bibinfo{person}{Richmond~Y Wong}, {and} \bibinfo{person}{Nick Merrill}.} \bibinfo{year}{2023}\natexlab{}.
\newblock \showarticletitle{Accidentally Evil: On Questionable Values in Smart Home Co-Design}. In \bibinfo{booktitle}{\emph{Proceedings of the 2023 CHI Conference on Human Factors in Computing Systems}}. \bibinfo{pages}{1--14}.
\newblock


\bibitem[Boyatzis(1998)]%
        {boyatzis1998transforming}
\bibfield{author}{\bibinfo{person}{Richard~E Boyatzis}.} \bibinfo{year}{1998}\natexlab{}.
\newblock \bibinfo{booktitle}{\emph{Transforming qualitative information: Thematic analysis and code development}}.
\newblock \bibinfo{publisher}{sage}.
\newblock


\bibitem[Brackenbury et~al\mbox{.}(2019)]%
        {brackenbury2019users}
\bibfield{author}{\bibinfo{person}{Will Brackenbury}, \bibinfo{person}{Abhimanyu Deora}, \bibinfo{person}{Jillian Ritchey}, \bibinfo{person}{Jason Vallee}, \bibinfo{person}{Weijia He}, \bibinfo{person}{Guan Wang}, \bibinfo{person}{Michael~L Littman}, {and} \bibinfo{person}{Blase Ur}.} \bibinfo{year}{2019}\natexlab{}.
\newblock \showarticletitle{How users interpret bugs in trigger-action programming}. In \bibinfo{booktitle}{\emph{Proceedings of the 2019 CHI conference on human factors in computing systems}}. \bibinfo{pages}{1--12}.
\newblock


\bibitem[Buchenau and Suri(2000)]%
        {buchenau2000experience}
\bibfield{author}{\bibinfo{person}{Marion Buchenau} {and} \bibinfo{person}{Jane~Fulton Suri}.} \bibinfo{year}{2000}\natexlab{}.
\newblock \showarticletitle{Experience prototyping}. In \bibinfo{booktitle}{\emph{Proceedings of the 3rd conference on Designing interactive systems: processes, practices, methods, and techniques}}. \bibinfo{pages}{424--433}.
\newblock


\bibitem[Chalhoub et~al\mbox{.}(2021)]%
        {chalhoub2021did}
\bibfield{author}{\bibinfo{person}{George Chalhoub}, \bibinfo{person}{Martin~J Kraemer}, \bibinfo{person}{Norbert Nthala}, {and} \bibinfo{person}{Ivan Flechais}.} \bibinfo{year}{2021}\natexlab{}.
\newblock \showarticletitle{“It did not give me an option to decline”: A Longitudinal Analysis of the User Experience of Security and Privacy in Smart Home Products}. In \bibinfo{booktitle}{\emph{Proceedings of the 2021 CHI Conference on Human Factors in Computing Systems}}. \bibinfo{pages}{1--16}.
\newblock


\bibitem[Cheng et~al\mbox{.}(2019)]%
        {cheng2019peekaboo}
\bibfield{author}{\bibinfo{person}{Yu-Ting Cheng}, \bibinfo{person}{Mathias Funk}, \bibinfo{person}{Wenn-Chieh Tsai}, {and} \bibinfo{person}{Lin-Lin Chen}.} \bibinfo{year}{2019}\natexlab{}.
\newblock \showarticletitle{Peekaboo Cam: Designing an observational camera for home ecologies concerning privacy}. In \bibinfo{booktitle}{\emph{Proceedings of the 2019 on Designing Interactive Systems Conference}}. \bibinfo{pages}{823--836}.
\newblock


\bibitem[Chiang et~al\mbox{.}(2020)]%
        {chiang2020exploring}
\bibfield{author}{\bibinfo{person}{Yi-Shyuan Chiang}, \bibinfo{person}{Ruei-Che Chang}, \bibinfo{person}{Yi-Lin Chuang}, \bibinfo{person}{Shih-Ya Chou}, \bibinfo{person}{Hao-Ping Lee}, \bibinfo{person}{I-Ju Lin}, \bibinfo{person}{Jian-Hua Jiang~Chen}, {and} \bibinfo{person}{Yung-Ju Chang}.} \bibinfo{year}{2020}\natexlab{}.
\newblock \showarticletitle{Exploring the design space of user-system communication for smart-home routine assistants}. In \bibinfo{booktitle}{\emph{Proceedings of the 2020 CHI Conference on Human Factors in Computing Systems}}. \bibinfo{pages}{1--14}.
\newblock


\bibitem[Cho et~al\mbox{.}(2023)]%
        {cho2023ai}
\bibfield{author}{\bibinfo{person}{Sungjae Cho}, \bibinfo{person}{Yoonsu Kim}, \bibinfo{person}{Jaewoong Jang}, {and} \bibinfo{person}{Inseok Hwang}.} \bibinfo{year}{2023}\natexlab{}.
\newblock \showarticletitle{AI-to-Human Actuation: Boosting Unmodified AI's Robustness by Proactively Inducing Favorable Human Sensing Conditions}.
\newblock \bibinfo{journal}{\emph{Proceedings of the ACM on Interactive, Mobile, Wearable and Ubiquitous Technologies}} \bibinfo{volume}{7}, \bibinfo{number}{1} (\bibinfo{year}{2023}), \bibinfo{pages}{1--32}.
\newblock


\bibitem[Coppers et~al\mbox{.}(2020)]%
        {coppers2020fortniot}
\bibfield{author}{\bibinfo{person}{Sven Coppers}, \bibinfo{person}{Davy Vanacken}, {and} \bibinfo{person}{Kris Luyten}.} \bibinfo{year}{2020}\natexlab{}.
\newblock \showarticletitle{Fortniot: Intelligible predictions to improve user understanding of smart home behavior}.
\newblock \bibinfo{journal}{\emph{Proceedings of the ACM on Interactive, Mobile, Wearable and Ubiquitous Technologies}} \bibinfo{volume}{4}, \bibinfo{number}{4} (\bibinfo{year}{2020}), \bibinfo{pages}{1--24}.
\newblock


\bibitem[Dai et~al\mbox{.}(2023)]%
        {dai2023effect}
\bibfield{author}{\bibinfo{person}{Jiaxin Dai}, \bibinfo{person}{Chao Zhang}, \bibinfo{person}{Dzmitry Aliakseyeu}, \bibinfo{person}{Samantha Peeters}, {and} \bibinfo{person}{Wijnand~A Ijsselsteijn}.} \bibinfo{year}{2023}\natexlab{}.
\newblock \showarticletitle{The Effect of Explanation Design on User Perception of Smart Home Lighting Systems: A Mixed-method Investigation}. In \bibinfo{booktitle}{\emph{Proceedings of the 2023 CHI Conference on Human Factors in Computing Systems}}. \bibinfo{pages}{1--14}.
\newblock


\bibitem[Das et~al\mbox{.}(2023)]%
        {das2023explainable}
\bibfield{author}{\bibinfo{person}{Devleena Das}, \bibinfo{person}{Yasutaka Nishimura}, \bibinfo{person}{Rajan~P Vivek}, \bibinfo{person}{Naoto Takeda}, \bibinfo{person}{Sean~T Fish}, \bibinfo{person}{Thomas Ploetz}, {and} \bibinfo{person}{Sonia Chernova}.} \bibinfo{year}{2023}\natexlab{}.
\newblock \showarticletitle{Explainable activity recognition for smart home systems}.
\newblock \bibinfo{journal}{\emph{ACM Transactions on Interactive Intelligent Systems}} \bibinfo{volume}{13}, \bibinfo{number}{2} (\bibinfo{year}{2023}), \bibinfo{pages}{1--39}.
\newblock


\bibitem[De~Ruyck et~al\mbox{.}(2019)]%
        {de2019user}
\bibfield{author}{\bibinfo{person}{Olivia De~Ruyck}, \bibinfo{person}{Peter Conradie}, \bibinfo{person}{Lieven De~Marez}, {and} \bibinfo{person}{Jelle Saldien}.} \bibinfo{year}{2019}\natexlab{}.
\newblock \showarticletitle{User needs in smart homes: changing needs according to life cycles and the impact on designing smart home solutions}. In \bibinfo{booktitle}{\emph{IFIP Conference on Human-Computer Interaction}}. Springer, \bibinfo{pages}{536--551}.
\newblock


\bibitem[Ehrenberg and Keinonen(2021)]%
        {ehrenberg2021technology}
\bibfield{author}{\bibinfo{person}{Nils Ehrenberg} {and} \bibinfo{person}{Turkka Keinonen}.} \bibinfo{year}{2021}\natexlab{}.
\newblock \showarticletitle{The technology is enemy for me at the moment: How smart home technologies assert control beyond intent}. In \bibinfo{booktitle}{\emph{Proceedings of the 2021 CHI Conference on Human Factors in Computing Systems}}. \bibinfo{pages}{1--11}.
\newblock


\bibitem[Garg and Cui(2022)]%
        {garg2022social}
\bibfield{author}{\bibinfo{person}{Radhika Garg} {and} \bibinfo{person}{Hua Cui}.} \bibinfo{year}{2022}\natexlab{}.
\newblock \showarticletitle{Social contexts, agency, and conflicts: Exploring critical aspects of design for future smart home technologies}.
\newblock \bibinfo{journal}{\emph{ACM Transactions on Computer-Human Interaction}} \bibinfo{volume}{29}, \bibinfo{number}{2} (\bibinfo{year}{2022}), \bibinfo{pages}{1--30}.
\newblock


\bibitem[Gonzalez~Penuela et~al\mbox{.}(2024)]%
        {gonzalez2024investigating}
\bibfield{author}{\bibinfo{person}{Ricardo~E Gonzalez~Penuela}, \bibinfo{person}{Jazmin Collins}, \bibinfo{person}{Cynthia Bennett}, {and} \bibinfo{person}{Shiri Azenkot}.} \bibinfo{year}{2024}\natexlab{}.
\newblock \showarticletitle{Investigating Use Cases of AI-Powered Scene Description Applications for Blind and Low Vision People}. In \bibinfo{booktitle}{\emph{Proceedings of the CHI Conference on Human Factors in Computing Systems}}. \bibinfo{pages}{1--21}.
\newblock


\bibitem[Harrington et~al\mbox{.}(2022)]%
        {harrington2022s}
\bibfield{author}{\bibinfo{person}{Christina~N Harrington}, \bibinfo{person}{Radhika Garg}, \bibinfo{person}{Amanda Woodward}, {and} \bibinfo{person}{Dimitri Williams}.} \bibinfo{year}{2022}\natexlab{}.
\newblock \showarticletitle{“It’s kind of like code-switching”: Black older adults’ experiences with a voice assistant for health information seeking}. In \bibinfo{booktitle}{\emph{Proceedings of the 2022 CHI Conference on Human Factors in Computing Systems}}. \bibinfo{pages}{1--15}.
\newblock


\bibitem[He et~al\mbox{.}(2019)]%
        {he2019smart}
\bibfield{author}{\bibinfo{person}{Weijia He}, \bibinfo{person}{Jesse Martinez}, \bibinfo{person}{Roshni Padhi}, \bibinfo{person}{Lefan Zhang}, {and} \bibinfo{person}{Blase Ur}.} \bibinfo{year}{2019}\natexlab{}.
\newblock \showarticletitle{When smart devices are stupid: negative experiences using home smart devices}. In \bibinfo{booktitle}{\emph{2019 IEEE Security and Privacy Workshops (SPW)}}. IEEE, \bibinfo{pages}{150--155}.
\newblock


\bibitem[Hu et~al\mbox{.}(2025)]%
        {hu2025elegnt}
\bibfield{author}{\bibinfo{person}{Yuhan Hu}, \bibinfo{person}{Peide Huang}, \bibinfo{person}{Mouli Sivapurapu}, {and} \bibinfo{person}{Jian Zhang}.} \bibinfo{year}{2025}\natexlab{}.
\newblock \showarticletitle{ELEGNT: Expressive and Functional Movement Design for Non-anthropomorphic Robot}.
\newblock \bibinfo{journal}{\emph{arXiv preprint arXiv:2501.12493}} (\bibinfo{year}{2025}).
\newblock


\bibitem[Huang and Cakmak(2015)]%
        {huang2015supporting}
\bibfield{author}{\bibinfo{person}{Justin Huang} {and} \bibinfo{person}{Maya Cakmak}.} \bibinfo{year}{2015}\natexlab{}.
\newblock \showarticletitle{Supporting mental model accuracy in trigger-action programming}. In \bibinfo{booktitle}{\emph{Proceedings of the 2015 acm international joint conference on pervasive and ubiquitous computing}}. \bibinfo{pages}{215--225}.
\newblock


\bibitem[Jakobi et~al\mbox{.}(2017)]%
        {jakobi2017catch}
\bibfield{author}{\bibinfo{person}{Timo Jakobi}, \bibinfo{person}{Corinna Ogonowski}, \bibinfo{person}{Nico Castelli}, \bibinfo{person}{Gunnar Stevens}, {and} \bibinfo{person}{Volker Wulf}.} \bibinfo{year}{2017}\natexlab{}.
\newblock \showarticletitle{The catch (es) with smart home: Experiences of a living lab field study}. In \bibinfo{booktitle}{\emph{Proceedings of the 2017 CHI Conference on Human Factors in Computing Systems}}. \bibinfo{pages}{1620--1633}.
\newblock


\bibitem[Jakobi et~al\mbox{.}(2018)]%
        {jakobi2018evolving}
\bibfield{author}{\bibinfo{person}{Timo Jakobi}, \bibinfo{person}{Gunnar Stevens}, \bibinfo{person}{Nico Castelli}, \bibinfo{person}{Corinna Ogonowski}, \bibinfo{person}{Florian Schaub}, \bibinfo{person}{Nils Vindice}, \bibinfo{person}{Dave Randall}, \bibinfo{person}{Peter Tolmie}, {and} \bibinfo{person}{Volker Wulf}.} \bibinfo{year}{2018}\natexlab{}.
\newblock \showarticletitle{Evolving needs in IoT control and accountability: A longitudinal study on smart home intelligibility}.
\newblock \bibinfo{journal}{\emph{Proceedings of the ACM on Interactive, Mobile, Wearable and Ubiquitous Technologies}} \bibinfo{volume}{2}, \bibinfo{number}{4} (\bibinfo{year}{2018}), \bibinfo{pages}{1--28}.
\newblock


\bibitem[Jensen et~al\mbox{.}(2018)]%
        {jensen2018designing}
\bibfield{author}{\bibinfo{person}{Rikke~Hagensby Jensen}, \bibinfo{person}{Yolande Strengers}, \bibinfo{person}{Jesper Kjeldskov}, \bibinfo{person}{Larissa Nicholls}, {and} \bibinfo{person}{Mikael~B Skov}.} \bibinfo{year}{2018}\natexlab{}.
\newblock \showarticletitle{Designing the desirable smart home: A study of household experiences and energy consumption impacts}. In \bibinfo{booktitle}{\emph{Proceedings of the 2018 CHI Conference on Human Factors in Computing Systems}}. \bibinfo{pages}{1--14}.
\newblock


\bibitem[Jin et~al\mbox{.}(2022)]%
        {jin2022exploring}
\bibfield{author}{\bibinfo{person}{Haojian Jin}, \bibinfo{person}{Boyuan Guo}, \bibinfo{person}{Rituparna Roychoudhury}, \bibinfo{person}{Yaxing Yao}, \bibinfo{person}{Swarun Kumar}, \bibinfo{person}{Yuvraj Agarwal}, {and} \bibinfo{person}{Jason~I Hong}.} \bibinfo{year}{2022}\natexlab{}.
\newblock \showarticletitle{Exploring the needs of users for supporting privacy-protective behaviors in smart homes}. In \bibinfo{booktitle}{\emph{Proceedings of the 2022 CHI Conference on Human Factors in Computing Systems}}. \bibinfo{pages}{1--19}.
\newblock


\bibitem[Kim and Nam(2022)]%
        {kim2022exploration}
\bibfield{author}{\bibinfo{person}{Chang-Min Kim} {and} \bibinfo{person}{Tek-Jin Nam}.} \bibinfo{year}{2022}\natexlab{}.
\newblock \showarticletitle{Exploration on Everyday Objects as an IoT Control Interface}. In \bibinfo{booktitle}{\emph{Designing Interactive Systems Conference}}. \bibinfo{pages}{1654--1668}.
\newblock


\bibitem[King et~al\mbox{.}(2024)]%
        {king2024sasha}
\bibfield{author}{\bibinfo{person}{Evan King}, \bibinfo{person}{Haoxiang Yu}, \bibinfo{person}{Sangsu Lee}, {and} \bibinfo{person}{Christine Julien}.} \bibinfo{year}{2024}\natexlab{}.
\newblock \showarticletitle{Sasha: creative goal-oriented reasoning in smart homes with large language models}.
\newblock \bibinfo{journal}{\emph{Proceedings of the ACM on Interactive, Mobile, Wearable and Ubiquitous Technologies}} \bibinfo{volume}{8}, \bibinfo{number}{1} (\bibinfo{year}{2024}), \bibinfo{pages}{1--38}.
\newblock


\bibitem[Klapperich et~al\mbox{.}(2020)]%
        {klapperich2020designing}
\bibfield{author}{\bibinfo{person}{Holger Klapperich}, \bibinfo{person}{Alarith Uhde}, {and} \bibinfo{person}{Marc Hassenzahl}.} \bibinfo{year}{2020}\natexlab{}.
\newblock \showarticletitle{Designing everyday automation with well-being in mind}.
\newblock \bibinfo{journal}{\emph{Personal and Ubiquitous Computing}} \bibinfo{volume}{24}, \bibinfo{number}{6} (\bibinfo{year}{2020}), \bibinfo{pages}{763--779}.
\newblock


\bibitem[Koshy et~al\mbox{.}(2021)]%
        {koshy2021we}
\bibfield{author}{\bibinfo{person}{Vinay Koshy}, \bibinfo{person}{Joon Sung~Sung Park}, \bibinfo{person}{Ti-Chung Cheng}, {and} \bibinfo{person}{Karrie Karahalios}.} \bibinfo{year}{2021}\natexlab{}.
\newblock \showarticletitle{“We Just Use What They Give Us”: Understanding Passenger User Perspectives in Smart Homes}. In \bibinfo{booktitle}{\emph{Proceedings of the 2021 CHI Conference on Human Factors in Computing Systems}}. \bibinfo{pages}{1--14}.
\newblock


\bibitem[Kurze et~al\mbox{.}(2020)]%
        {kurze2020guess}
\bibfield{author}{\bibinfo{person}{Albrecht Kurze}, \bibinfo{person}{Andreas Bischof}, \bibinfo{person}{S{\"o}ren Totzauer}, \bibinfo{person}{Michael Storz}, \bibinfo{person}{Maximilian Eibl}, \bibinfo{person}{Margot Brereton}, {and} \bibinfo{person}{Arne Berger}.} \bibinfo{year}{2020}\natexlab{}.
\newblock \showarticletitle{Guess the data: Data work to understand how people make sense of and use simple sensor data from homes}. In \bibinfo{booktitle}{\emph{Proceedings of the 2020 CHI conference on human factors in computing systems}}. \bibinfo{pages}{1--12}.
\newblock


\bibitem[Lee et~al\mbox{.}(2024)]%
        {lee2024gazepointar}
\bibfield{author}{\bibinfo{person}{Jaewook Lee}, \bibinfo{person}{Jun Wang}, \bibinfo{person}{Elizabeth Brown}, \bibinfo{person}{Liam Chu}, \bibinfo{person}{Sebastian~S Rodriguez}, {and} \bibinfo{person}{Jon~E Froehlich}.} \bibinfo{year}{2024}\natexlab{}.
\newblock \showarticletitle{GazePointAR: A Context-Aware Multimodal Voice Assistant for Pronoun Disambiguation in Wearable Augmented Reality}. In \bibinfo{booktitle}{\emph{Proceedings of the CHI Conference on Human Factors in Computing Systems}}. \bibinfo{pages}{1--20}.
\newblock


\bibitem[Lee and Qiufan(2021)]%
        {lee2021ai}
\bibfield{author}{\bibinfo{person}{Kai-Fu Lee} {and} \bibinfo{person}{Chen Qiufan}.} \bibinfo{year}{2021}\natexlab{}.
\newblock \bibinfo{booktitle}{\emph{AI 2041: Ten visions for our future}}.
\newblock \bibinfo{publisher}{Crown Currency}.
\newblock


\bibitem[Lee and Lim(2024)]%
        {lee2024we}
\bibfield{author}{\bibinfo{person}{Yoomi Lee} {and} \bibinfo{person}{Youn-kyung Lim}.} \bibinfo{year}{2024}\natexlab{}.
\newblock \showarticletitle{How We Use Together: Coordinating Individual Preferences for Using Shared Devices at Home}. In \bibinfo{booktitle}{\emph{Proceedings of the 2024 ACM Designing Interactive Systems Conference}}. \bibinfo{pages}{3407--3418}.
\newblock


\bibitem[Li et~al\mbox{.}(2024b)]%
        {li2024conversational}
\bibfield{author}{\bibinfo{person}{Brenna Li}, \bibinfo{person}{Amy Wang}, \bibinfo{person}{Patricia Strachan}, \bibinfo{person}{Julie~Anne S{\'e}guin}, \bibinfo{person}{Sami Lachgar}, \bibinfo{person}{Karyn~C Schroeder}, \bibinfo{person}{Mathias~S Fleck}, \bibinfo{person}{Renee Wong}, \bibinfo{person}{Alan Karthikesalingam}, \bibinfo{person}{Vivek Natarajan}, {et~al\mbox{.}}} \bibinfo{year}{2024}\natexlab{b}.
\newblock \showarticletitle{Conversational AI in health: Design considerations from a Wizard-of-Oz dermatology case study with users, clinicians and a medical LLM}. In \bibinfo{booktitle}{\emph{Extended Abstracts of the CHI Conference on Human Factors in Computing Systems}}. \bibinfo{pages}{1--10}.
\newblock


\bibitem[Li et~al\mbox{.}(2019)]%
        {li2019fmt}
\bibfield{author}{\bibinfo{person}{Franklin~Mingzhe Li}, \bibinfo{person}{Di~Laura Chen}, \bibinfo{person}{Mingming Fan}, {and} \bibinfo{person}{Khai~N Truong}.} \bibinfo{year}{2019}\natexlab{}.
\newblock \showarticletitle{FMT: A wearable camera-based object tracking memory aid for older adults}.
\newblock \bibinfo{journal}{\emph{Proceedings of the ACM on Interactive, Mobile, Wearable and Ubiquitous Technologies}} \bibinfo{volume}{3}, \bibinfo{number}{3} (\bibinfo{year}{2019}), \bibinfo{pages}{1--25}.
\newblock


\bibitem[Li et~al\mbox{.}(2024a)]%
        {li2024contextual}
\bibfield{author}{\bibinfo{person}{Franklin~Mingzhe Li}, \bibinfo{person}{Michael~Xieyang Liu}, \bibinfo{person}{Shaun~K Kane}, {and} \bibinfo{person}{Patrick Carrington}.} \bibinfo{year}{2024}\natexlab{a}.
\newblock \showarticletitle{A Contextual Inquiry of People with Vision Impairments in Cooking}. In \bibinfo{booktitle}{\emph{Proceedings of the CHI Conference on Human Factors in Computing Systems}}. \bibinfo{pages}{1--14}.
\newblock


\bibitem[Liu et~al\mbox{.}(2023)]%
        {liu2023understanding}
\bibfield{author}{\bibinfo{person}{Xiaoyi Liu}, \bibinfo{person}{Yingtian Shi}, \bibinfo{person}{Chun Yu}, \bibinfo{person}{Cheng Gao}, \bibinfo{person}{Tianao Yang}, \bibinfo{person}{Chen Liang}, {and} \bibinfo{person}{Yuanchun Shi}.} \bibinfo{year}{2023}\natexlab{}.
\newblock \showarticletitle{Understanding In-Situ Programming for Smart Home Automation}.
\newblock \bibinfo{journal}{\emph{Proceedings of the ACM on Interactive, Mobile, Wearable and Ubiquitous Technologies}} \bibinfo{volume}{7}, \bibinfo{number}{2} (\bibinfo{year}{2023}), \bibinfo{pages}{1--31}.
\newblock


\bibitem[Liu et~al\mbox{.}(2024)]%
        {liu2024unblind}
\bibfield{author}{\bibinfo{person}{Zhe Liu}, \bibinfo{person}{Chunyang Chen}, \bibinfo{person}{Junjie Wang}, \bibinfo{person}{Mengzhuo Chen}, \bibinfo{person}{Boyu Wu}, \bibinfo{person}{Yuekai Huang}, \bibinfo{person}{Jun Hu}, {and} \bibinfo{person}{Qing Wang}.} \bibinfo{year}{2024}\natexlab{}.
\newblock \showarticletitle{Unblind Text Inputs: Predicting Hint-text of Text Input in Mobile Apps via LLM}. In \bibinfo{booktitle}{\emph{Proceedings of the CHI Conference on Human Factors in Computing Systems}}. \bibinfo{pages}{1--20}.
\newblock


\bibitem[Moore et~al\mbox{.}(2018)]%
        {moore2018managing}
\bibfield{author}{\bibinfo{person}{Jimmy Moore}, \bibinfo{person}{Pascal Goffin}, \bibinfo{person}{Miriah Meyer}, \bibinfo{person}{Philip Lundrigan}, \bibinfo{person}{Neal Patwari}, \bibinfo{person}{Katherine Sward}, {and} \bibinfo{person}{Jason Wiese}.} \bibinfo{year}{2018}\natexlab{}.
\newblock \showarticletitle{Managing in-home environments through sensing, annotating, and visualizing air quality data}.
\newblock \bibinfo{journal}{\emph{Proceedings of the ACM on interactive, mobile, wearable and ubiquitous technologies}} \bibinfo{volume}{2}, \bibinfo{number}{3} (\bibinfo{year}{2018}), \bibinfo{pages}{1--28}.
\newblock


\bibitem[Ning et~al\mbox{.}(2024)]%
        {ning2024spica}
\bibfield{author}{\bibinfo{person}{Zheng Ning}, \bibinfo{person}{Brianna~L Wimer}, \bibinfo{person}{Kaiwen Jiang}, \bibinfo{person}{Keyi Chen}, \bibinfo{person}{Jerrick Ban}, \bibinfo{person}{Yapeng Tian}, \bibinfo{person}{Yuhang Zhao}, {and} \bibinfo{person}{Toby Jia-Jun Li}.} \bibinfo{year}{2024}\natexlab{}.
\newblock \showarticletitle{SPICA: Interactive Video Content Exploration through Augmented Audio Descriptions for Blind or Low-Vision Viewers}. In \bibinfo{booktitle}{\emph{Proceedings of the CHI Conference on Human Factors in Computing Systems}}. \bibinfo{pages}{1--18}.
\newblock


\bibitem[Pierce(2019)]%
        {pierce2019smart}
\bibfield{author}{\bibinfo{person}{James Pierce}.} \bibinfo{year}{2019}\natexlab{}.
\newblock \showarticletitle{Smart home security cameras and shifting lines of creepiness: A design-led inquiry}. In \bibinfo{booktitle}{\emph{Proceedings of the 2019 CHI Conference on Human Factors in Computing Systems}}. \bibinfo{pages}{1--14}.
\newblock


\bibitem[Pierce et~al\mbox{.}(2022)]%
        {pierce2022addressing}
\bibfield{author}{\bibinfo{person}{James Pierce}, \bibinfo{person}{Claire Weizenegger}, \bibinfo{person}{Parag Nandi}, \bibinfo{person}{Isha Agarwal}, \bibinfo{person}{Gwenna Gram}, \bibinfo{person}{Jade Hurrle}, \bibinfo{person}{Hannah Liao}, \bibinfo{person}{Betty Lo}, \bibinfo{person}{Aaron Park}, \bibinfo{person}{Aivy Phan}, {et~al\mbox{.}}} \bibinfo{year}{2022}\natexlab{}.
\newblock \showarticletitle{Addressing Adjacent Actor Privacy: Designing for Bystanders, Co-Users, and Surveilled Subjects of Smart Home Cameras}. In \bibinfo{booktitle}{\emph{Designing Interactive Systems Conference}}. \bibinfo{pages}{26--40}.
\newblock


\bibitem[Reisinger et~al\mbox{.}(2017)]%
        {reisinger2017visual}
\bibfield{author}{\bibinfo{person}{Michaela~R Reisinger}, \bibinfo{person}{Johann Schrammel}, {and} \bibinfo{person}{Peter Fr{\"o}hlich}.} \bibinfo{year}{2017}\natexlab{}.
\newblock \showarticletitle{Visual languages for smart spaces: End-user programming between data-flow and form-filling}. In \bibinfo{booktitle}{\emph{2017 IEEE Symposium on Visual Languages and Human-Centric Computing (VL/HCC)}}. IEEE, \bibinfo{pages}{165--169}.
\newblock


\bibitem[Ripley and Politzer(2010)]%
        {ripley2010vision}
\bibfield{author}{\bibinfo{person}{David~L Ripley} {and} \bibinfo{person}{Thomas Politzer}.} \bibinfo{year}{2010}\natexlab{}.
\newblock \showarticletitle{Vision disturbance after TBI}.
\newblock \bibinfo{journal}{\emph{NeuroRehabilitation}} \bibinfo{volume}{27}, \bibinfo{number}{3} (\bibinfo{year}{2010}), \bibinfo{pages}{215}.
\newblock


\bibitem[Sarkar(2024)]%
        {challenge}
\bibfield{author}{\bibinfo{person}{Advait Sarkar}.} \bibinfo{year}{2024}\natexlab{}.
\newblock
\newblock
\urldef\tempurl%
\url{https://cacm.acm.org/opinion/ai-should-challenge-not-obey/}
\showURL{%
\tempurl}


\bibitem[Shalawadi et~al\mbox{.}(2024)]%
        {shalawadi2024manual}
\bibfield{author}{\bibinfo{person}{Sujay Shalawadi}, \bibinfo{person}{Christopher Getschmann}, \bibinfo{person}{Niels van Berkel}, {and} \bibinfo{person}{Florian Echtler}.} \bibinfo{year}{2024}\natexlab{}.
\newblock \showarticletitle{Manual, Hybrid, and Automatic Privacy Covers for Smart Home Cameras}. In \bibinfo{booktitle}{\emph{Proceedings of the 2024 ACM Designing Interactive Systems Conference}}. \bibinfo{pages}{3453--3470}.
\newblock


\bibitem[Strengers et~al\mbox{.}(2022)]%
        {strengers2022isn}
\bibfield{author}{\bibinfo{person}{Yolande Strengers}, \bibinfo{person}{Melisa Duque}, \bibinfo{person}{Michael Mortimer}, \bibinfo{person}{Sarah Pink}, \bibinfo{person}{Rex Martin}, \bibinfo{person}{Larissa Nicholls}, \bibinfo{person}{Ben Horan}, \bibinfo{person}{Alicia Eugene}, {and} \bibinfo{person}{Sue Thomson}.} \bibinfo{year}{2022}\natexlab{}.
\newblock \showarticletitle{“Isn't this Marvelous” Supporting Older Adults’ Wellbeing with Smart Home Devices Through Curiosity, Play and Experimentation}. In \bibinfo{booktitle}{\emph{Proceedings of the 2022 ACM Designing Interactive Systems Conference}}. \bibinfo{pages}{707--725}.
\newblock


\bibitem[Strengers et~al\mbox{.}(2019)]%
        {strengers2019protection}
\bibfield{author}{\bibinfo{person}{Yolande Strengers}, \bibinfo{person}{Jenny Kennedy}, \bibinfo{person}{Paula Arcari}, \bibinfo{person}{Larissa Nicholls}, {and} \bibinfo{person}{Melissa Gregg}.} \bibinfo{year}{2019}\natexlab{}.
\newblock \showarticletitle{Protection, productivity and pleasure in the smart home: Emerging expectations and gendered insights from Australian early adopters}. In \bibinfo{booktitle}{\emph{Proceedings of the 2019 CHI conference on human factors in computing systems}}. \bibinfo{pages}{1--13}.
\newblock


\bibitem[Su et~al\mbox{.}(2024)]%
        {su2024rassar}
\bibfield{author}{\bibinfo{person}{Xia Su}, \bibinfo{person}{Han Zhang}, \bibinfo{person}{Kaiming Cheng}, \bibinfo{person}{Jaewook Lee}, \bibinfo{person}{Qiaochu Liu}, \bibinfo{person}{Wyatt Olson}, {and} \bibinfo{person}{Jon~E Froehlich}.} \bibinfo{year}{2024}\natexlab{}.
\newblock \showarticletitle{RASSAR: Room Accessibility and Safety Scanning in Augmented Reality}. In \bibinfo{booktitle}{\emph{Proceedings of the CHI Conference on Human Factors in Computing Systems}}. \bibinfo{pages}{1--17}.
\newblock


\bibitem[Szymanski et~al\mbox{.}(2024)]%
        {szymanski2024integrating}
\bibfield{author}{\bibinfo{person}{Annalisa Szymanski}, \bibinfo{person}{Brianna~L Wimer}, \bibinfo{person}{Oghenemaro Anuyah}, \bibinfo{person}{Heather~A Eicher-Miller}, {and} \bibinfo{person}{Ronald~A Metoyer}.} \bibinfo{year}{2024}\natexlab{}.
\newblock \showarticletitle{Integrating Expertise in LLMs: Crafting a Customized Nutrition Assistant with Refined Template Instructions}. In \bibinfo{booktitle}{\emph{Proceedings of the CHI Conference on Human Factors in Computing Systems}}. \bibinfo{pages}{1--22}.
\newblock


\bibitem[Tao and Vyas(2021)]%
        {tao2021diy}
\bibfield{author}{\bibinfo{person}{Hongyi Tao} {and} \bibinfo{person}{Dhaval Vyas}.} \bibinfo{year}{2021}\natexlab{}.
\newblock \showarticletitle{DIY Homes: Placemaking in Rural Eco-Homes}. In \bibinfo{booktitle}{\emph{IFIP Conference on Human-Computer Interaction}}. Springer, \bibinfo{pages}{343--364}.
\newblock


\bibitem[Team et~al\mbox{.}(2023)]%
        {team2023gemini}
\bibfield{author}{\bibinfo{person}{Gemini Team}, \bibinfo{person}{Rohan Anil}, \bibinfo{person}{Sebastian Borgeaud}, \bibinfo{person}{Yonghui Wu}, \bibinfo{person}{Jean-Baptiste Alayrac}, \bibinfo{person}{Jiahui Yu}, \bibinfo{person}{Radu Soricut}, \bibinfo{person}{Johan Schalkwyk}, \bibinfo{person}{Andrew~M Dai}, \bibinfo{person}{Anja Hauth}, {et~al\mbox{.}}} \bibinfo{year}{2023}\natexlab{}.
\newblock \showarticletitle{Gemini: a family of highly capable multimodal models}.
\newblock \bibinfo{journal}{\emph{arXiv preprint arXiv:2312.11805}} (\bibinfo{year}{2023}).
\newblock


\bibitem[Ur et~al\mbox{.}(2014)]%
        {ur2014practical}
\bibfield{author}{\bibinfo{person}{Blase Ur}, \bibinfo{person}{Elyse McManus}, \bibinfo{person}{Melwyn Pak Yong~Ho}, {and} \bibinfo{person}{Michael~L Littman}.} \bibinfo{year}{2014}\natexlab{}.
\newblock \showarticletitle{Practical trigger-action programming in the smart home}. In \bibinfo{booktitle}{\emph{Proceedings of the SIGCHI conference on human factors in computing systems}}. \bibinfo{pages}{803--812}.
\newblock


\bibitem[Urquhart and Fern{\'a}ndez(2013)]%
        {urquhart2013using}
\bibfield{author}{\bibinfo{person}{Cathy Urquhart} {and} \bibinfo{person}{Walter Fern{\'a}ndez}.} \bibinfo{year}{2013}\natexlab{}.
\newblock \showarticletitle{Using grounded theory method in information systems: The researcher as blank slate and other myths}.
\newblock \bibinfo{journal}{\emph{Journal of Information Technology}}  \bibinfo{volume}{28} (\bibinfo{year}{2013}), \bibinfo{pages}{224--236}.
\newblock


\bibitem[Venugopalan et~al\mbox{.}(2024)]%
        {venugopalan2024aragorn}
\bibfield{author}{\bibinfo{person}{Hari Venugopalan}, \bibinfo{person}{Zainul~Abi Din}, \bibinfo{person}{Trevor Carpenter}, \bibinfo{person}{Jason Lowe-Power}, \bibinfo{person}{Samuel~T King}, {and} \bibinfo{person}{Zubair Shafiq}.} \bibinfo{year}{2024}\natexlab{}.
\newblock \showarticletitle{Aragorn: A Privacy-Enhancing System for Mobile Cameras}.
\newblock \bibinfo{journal}{\emph{Proceedings of the ACM on Interactive, Mobile, Wearable and Ubiquitous Technologies}} \bibinfo{volume}{7}, \bibinfo{number}{4} (\bibinfo{year}{2024}), \bibinfo{pages}{1--31}.
\newblock


\bibitem[Verweij(2019)]%
        {verweij2019exploring}
\bibfield{author}{\bibinfo{person}{David Verweij}.} \bibinfo{year}{2019}\natexlab{}.
\newblock \showarticletitle{Exploring Future IoT for Families through End User Development: Applying Do-It-Together Practises to Reveal Family Dynamics in Technology Adoption}. In \bibinfo{booktitle}{\emph{Extended Abstracts of the 2019 CHI Conference on Human Factors in Computing Systems}}. \bibinfo{pages}{1--4}.
\newblock


\bibitem[Wang et~al\mbox{.}(2024)]%
        {wang2024pepperpose}
\bibfield{author}{\bibinfo{person}{Chongyang Wang}, \bibinfo{person}{Siqi Zheng}, \bibinfo{person}{Lingxiao Zhong}, \bibinfo{person}{Chun Yu}, \bibinfo{person}{Chen Liang}, \bibinfo{person}{Yuntao Wang}, \bibinfo{person}{Yuan Gao}, \bibinfo{person}{Tin~Lun Lam}, {and} \bibinfo{person}{Yuanchun Shi}.} \bibinfo{year}{2024}\natexlab{}.
\newblock \showarticletitle{PepperPose: Full-Body Pose Estimation with a Companion Robot}. In \bibinfo{booktitle}{\emph{Proceedings of the CHI Conference on Human Factors in Computing Systems}}. \bibinfo{pages}{1--16}.
\newblock


\bibitem[Wen et~al\mbox{.}(2024)]%
        {wen2024find}
\bibfield{author}{\bibinfo{person}{Linda~Yilin Wen}, \bibinfo{person}{Cecily Morrison}, \bibinfo{person}{Martin Grayson}, \bibinfo{person}{Rita~Faia Marques}, \bibinfo{person}{Daniela Massiceti}, \bibinfo{person}{Camilla Longden}, {and} \bibinfo{person}{Edward Cutrell}.} \bibinfo{year}{2024}\natexlab{}.
\newblock \showarticletitle{Find My Things: Personalized Accessibility through Teachable AI for People who are Blind or Low Vision}. In \bibinfo{booktitle}{\emph{Extended Abstracts of the CHI Conference on Human Factors in Computing Systems}}. \bibinfo{pages}{1--6}.
\newblock


\bibitem[Williams et~al\mbox{.}(2019)]%
        {williams2019understanding}
\bibfield{author}{\bibinfo{person}{Kristin Williams}, \bibinfo{person}{Rajitha Pulivarthy}, \bibinfo{person}{Scott~E Hudson}, {and} \bibinfo{person}{Jessica Hammer}.} \bibinfo{year}{2019}\natexlab{}.
\newblock \showarticletitle{Understanding family collaboration around lightweight modification of everyday objects in the home}.
\newblock \bibinfo{journal}{\emph{Proceedings of the ACM on Human-Computer Interaction}} \bibinfo{volume}{3}, \bibinfo{number}{CSCW} (\bibinfo{year}{2019}), \bibinfo{pages}{1--24}.
\newblock


\bibitem[Williams et~al\mbox{.}(2020)]%
        {williams2020upcycled}
\bibfield{author}{\bibinfo{person}{Kristin Williams}, \bibinfo{person}{Rajitha Pulivarthy}, \bibinfo{person}{Scott~E Hudson}, {and} \bibinfo{person}{Jessica Hammer}.} \bibinfo{year}{2020}\natexlab{}.
\newblock \showarticletitle{The Upcycled Home: Removing barriers to lightweight modification of the home's everyday objects}. In \bibinfo{booktitle}{\emph{Proceedings of the 2020 CHI Conference on Human Factors in Computing Systems}}. \bibinfo{pages}{1--13}.
\newblock


\bibitem[Woo and Lim(2015)]%
        {woo2015user}
\bibfield{author}{\bibinfo{person}{Jong-bum Woo} {and} \bibinfo{person}{Youn-kyung Lim}.} \bibinfo{year}{2015}\natexlab{}.
\newblock \showarticletitle{User experience in do-it-yourself-style smart homes}. In \bibinfo{booktitle}{\emph{Proceedings of the 2015 ACM international joint conference on pervasive and ubiquitous computing}}. \bibinfo{pages}{779--790}.
\newblock


\bibitem[Wo{\'z}niak et~al\mbox{.}(2023)]%
        {wozniak2023inhabiting}
\bibfield{author}{\bibinfo{person}{Miko{\l}aj~P Wo{\'z}niak}, \bibinfo{person}{Sarah V{\"o}ge}, \bibinfo{person}{Ronja Kr{\"u}ger}, \bibinfo{person}{Heiko M{\"u}ller}, \bibinfo{person}{Marion Koelle}, {and} \bibinfo{person}{Susanne Boll}.} \bibinfo{year}{2023}\natexlab{}.
\newblock \showarticletitle{Inhabiting interconnected spaces: How users shape and appropriate their smart home ecosystems}. In \bibinfo{booktitle}{\emph{Proceedings of the 2023 CHI Conference on Human Factors in Computing Systems}}. \bibinfo{pages}{1--18}.
\newblock


\bibitem[Xu et~al\mbox{.}(2024)]%
        {xu2024dipa2}
\bibfield{author}{\bibinfo{person}{Anran Xu}, \bibinfo{person}{Zhongyi Zhou}, \bibinfo{person}{Kakeru Miyazaki}, \bibinfo{person}{Ryo Yoshikawa}, \bibinfo{person}{Simo Hosio}, {and} \bibinfo{person}{Koji Yatani}.} \bibinfo{year}{2024}\natexlab{}.
\newblock \showarticletitle{DIPA2: An Image Dataset with Cross-cultural Privacy Perception Annotations}.
\newblock \bibinfo{journal}{\emph{Proceedings of the ACM on Interactive, Mobile, Wearable and Ubiquitous Technologies}} \bibinfo{volume}{7}, \bibinfo{number}{4} (\bibinfo{year}{2024}), \bibinfo{pages}{1--30}.
\newblock


\bibitem[Xue et~al\mbox{.}(2024)]%
        {xue2024should}
\bibfield{author}{\bibinfo{person}{Xiao Xue}, \bibinfo{person}{Xinyang Li}, \bibinfo{person}{Boyang Jia}, \bibinfo{person}{Jiachen Du}, {and} \bibinfo{person}{Xinyi Fu}.} \bibinfo{year}{2024}\natexlab{}.
\newblock \showarticletitle{Who Should Hold Control? Rethinking Empowerment in Home Automation among Cohabitants through the Lens of Co-Design}. In \bibinfo{booktitle}{\emph{Proceedings of the CHI Conference on Human Factors in Computing Systems}}. \bibinfo{pages}{1--19}.
\newblock


\bibitem[Yeo et~al\mbox{.}(2023)]%
        {yeo2023omnisense}
\bibfield{author}{\bibinfo{person}{Hui-Shyong Yeo}, \bibinfo{person}{Erwin Wu}, \bibinfo{person}{Daehwa Kim}, \bibinfo{person}{Juyoung Lee}, \bibinfo{person}{Hyung-il Kim}, \bibinfo{person}{Seo~Young Oh}, \bibinfo{person}{Luna Takagi}, \bibinfo{person}{Woontack Woo}, \bibinfo{person}{Hideki Koike}, {and} \bibinfo{person}{Aaron~John Quigley}.} \bibinfo{year}{2023}\natexlab{}.
\newblock \showarticletitle{OmniSense: Exploring Novel Input Sensing and Interaction Techniques on Mobile Device with an Omni-Directional Camera}. In \bibinfo{booktitle}{\emph{Proceedings of the 2023 CHI Conference on Human Factors in Computing Systems}}. \bibinfo{pages}{1--18}.
\newblock


\bibitem[Yildirim et~al\mbox{.}(2024)]%
        {yildirim2024multimodal}
\bibfield{author}{\bibinfo{person}{Nur Yildirim}, \bibinfo{person}{Hannah Richardson}, \bibinfo{person}{Maria~Teodora Wetscherek}, \bibinfo{person}{Junaid Bajwa}, \bibinfo{person}{Joseph Jacob}, \bibinfo{person}{Mark~Ames Pinnock}, \bibinfo{person}{Stephen Harris}, \bibinfo{person}{Daniel Coelho De~Castro}, \bibinfo{person}{Shruthi Bannur}, \bibinfo{person}{Stephanie Hyland}, {et~al\mbox{.}}} \bibinfo{year}{2024}\natexlab{}.
\newblock \showarticletitle{Multimodal healthcare AI: identifying and designing clinically relevant vision-language applications for radiology}. In \bibinfo{booktitle}{\emph{Proceedings of the CHI Conference on Human Factors in Computing Systems}}. \bibinfo{pages}{1--22}.
\newblock


\bibitem[Yu et~al\mbox{.}(2021)]%
        {yu2021analysis}
\bibfield{author}{\bibinfo{person}{Haoxiang Yu}, \bibinfo{person}{Jie Hua}, {and} \bibinfo{person}{Christine Julien}.} \bibinfo{year}{2021}\natexlab{}.
\newblock \showarticletitle{Analysis of ifttt recipes to study how humans use internet-of-things (iot) devices}. In \bibinfo{booktitle}{\emph{Proceedings of the 19th ACM conference on embedded networked sensor systems}}. \bibinfo{pages}{537--541}.
\newblock


\bibitem[Yun and Lim(2023)]%
        {yun2023potential}
\bibfield{author}{\bibinfo{person}{Sojeong Yun} {and} \bibinfo{person}{Youn-Kyung Lim}.} \bibinfo{year}{2023}\natexlab{}.
\newblock \showarticletitle{Potential and Challenges of DIY Smart Homes with an ML-intensive Camera Sensor}. In \bibinfo{booktitle}{\emph{Proceedings of the 2023 CHI Conference on Human Factors in Computing Systems}}. \bibinfo{pages}{1--19}.
\newblock


\bibitem[Yun and Lim(2024)]%
        {yun2024toolkit}
\bibfield{author}{\bibinfo{person}{Sojeong Yun} {and} \bibinfo{person}{Youn-kyung Lim}.} \bibinfo{year}{2024}\natexlab{}.
\newblock \showarticletitle{Toolkit Design for Building Camera Sensor-Driven DIY Smart Homes}. In \bibinfo{booktitle}{\emph{Companion of the 2024 on ACM International Joint Conference on Pervasive and Ubiquitous Computing}}. \bibinfo{pages}{256--261}.
\newblock


\bibitem[Zaidi et~al\mbox{.}(2023)]%
        {zaidi2023user}
\bibfield{author}{\bibinfo{person}{Ali Zaidi}, \bibinfo{person}{Rui Yang}, \bibinfo{person}{Vinay Koshy}, \bibinfo{person}{Camille Cobb}, \bibinfo{person}{Indranil Gupta}, {and} \bibinfo{person}{Karrie Karahalios}.} \bibinfo{year}{2023}\natexlab{}.
\newblock \showarticletitle{A User-Centric Evaluation of Smart Home Resolution Approaches for Conflicts Between Routines}.
\newblock \bibinfo{journal}{\emph{Proceedings of the ACM on Interactive, Mobile, Wearable and Ubiquitous Technologies}} \bibinfo{volume}{7}, \bibinfo{number}{1} (\bibinfo{year}{2023}), \bibinfo{pages}{1--35}.
\newblock


\bibitem[Zhang et~al\mbox{.}(2023)]%
        {zhang2023helping}
\bibfield{author}{\bibinfo{person}{Lefan Zhang}, \bibinfo{person}{Cyrus Zhou}, \bibinfo{person}{Michael~L Littman}, \bibinfo{person}{Blase Ur}, {and} \bibinfo{person}{Shan Lu}.} \bibinfo{year}{2023}\natexlab{}.
\newblock \showarticletitle{Helping Users Debug Trigger-Action Programs}.
\newblock \bibinfo{journal}{\emph{Proceedings of the ACM on Interactive, Mobile, Wearable and Ubiquitous Technologies}} \bibinfo{volume}{6}, \bibinfo{number}{4} (\bibinfo{year}{2023}), \bibinfo{pages}{1--32}.
\newblock


\bibitem[Zhao et~al\mbox{.}(2021)]%
        {zhao2021understanding}
\bibfield{author}{\bibinfo{person}{Valerie Zhao}, \bibinfo{person}{Lefan Zhang}, \bibinfo{person}{Bo Wang}, \bibinfo{person}{Michael~L Littman}, \bibinfo{person}{Shan Lu}, {and} \bibinfo{person}{Blase Ur}.} \bibinfo{year}{2021}\natexlab{}.
\newblock \showarticletitle{Understanding trigger-action programs through novel visualizations of program differences}. In \bibinfo{booktitle}{\emph{Proceedings of the 2021 CHI Conference on Human Factors in Computing Systems}}. \bibinfo{pages}{1--17}.
\newblock


\end{thebibliography}



\begin{table*}[]
\caption{List of features created by participants} 
\begin{tabular}{lll}
\toprule[1pt]
\textbf{\begin{tabular}[c]{@{}l@{}}Three\\ Roles\end{tabular}} & \textbf{\begin{tabular}[c]{@{}l@{}}Purpose of\\ Developed Features\end{tabular}} & \textbf{Specific Features Developed by Participants} \\ \midrule[1pt]
\multirow{3}{*}{\begin{tabular}[c]{@{}l@{}}Auto-\\ monitoring\end{tabular}} & \begin{tabular}[c]{@{}l@{}}To help users better\\ recognize the current\\ state of their home\end{tabular} & \begin{tabular}[c]{@{}l@{}}Alerts for potentially dangerous cooking moments (P1,2),\\ left-on energy-wasting appliances (P1,8),\\ areas requiring cleaning or tidying (P1,2,3,5,6,7,10,12),\\ plants needing water (P5), full laundry baskets (P5,6),\\ a full sink of dishes (P5), the location of desired items (P8),\\ and pre-departure safety checks (P10)\end{tabular} \\ \cline{2-3} 
 & \begin{tabular}[c]{@{}l@{}}To help users\\ recognize their\\ habitual behaviors\end{tabular} & \begin{tabular}[c]{@{}l@{}}Alerts for posture correction while sitting (P4), nail-biting behavior (P10),\\ music playback during heightened emotions while working remotely (P3),\\ 30-minute mobile gaming alerts (P10), missed supplement intake alerts (P4,9),\\ alerts for exceeding recommended times\\ for exercise, gaming, or sleep (P1,10),\\ recommendations for meditation or exercise videos\\if lying in bed for over 2 hours (P10),\\ blinking light near the fridge for insufficient daily water intake (P4),\\ alerts if teeth are not brushed within 20 minutes of eating (P4),\\ alerts if children do not organize their bags after returning home (P2),\\ alerts if children stay awake past bedtime (P2), and a daily summary of activities (P6)\end{tabular} \\ \cline{2-3} 
 & \begin{tabular}[c]{@{}l@{}}To help users\\ recognize situations\\ requiring support\\ for family members\end{tabular} & \begin{tabular}[c]{@{}l@{}}Alerts if a baby climbs stairs alone (P6), alerts if a child falls (P5),\\ alerts when a baby wakes up or falls asleep (P6),\\ alerts if a child uses a desk without turning on the light (P2),\\ alerts if a pet shows anxiety symptoms (P11),\\ temperature adjustment when a cold posture is detected (P10),\\ activation of do-not-disturb mode when a family member is working remotely (P12),\\ and lighting and music adjustments based on family activities (P2)\end{tabular} \\ \midrule[0.5pt]
\multirow{2}{*}{Assistant} & \begin{tabular}[c]{@{}l@{}}To support more\\ precise and sensitive\\ judgment\\ than humans\end{tabular} & \begin{tabular}[c]{@{}l@{}}Assistance in detecting stains on clothes (P1),\\ guidance on specific areas requiring cleaning (P1,4),\\alerts for misplaced fridge items (P7),\\ recording unusual facial expressions of family members (P3),\\ notifying changes in plant color or signs of poor health (P5),\\ guidance on replenishing low-stocked alcoholic beverages (P7),\\ assistance in determining appropriate caregiving actions (P7,8),\\ alerts for less fresh cooking ingredients (P12),\\ notifications for expiring food items in the fridge (P1,5)\end{tabular} \\ \cline{2-3} 
 & \begin{tabular}[c]{@{}l@{}}To support an ideal\\ daily activity with\\ helpful information\end{tabular} & \begin{tabular}[c]{@{}l@{}}Nutritional guidance when taking out unhealthy foods from the fridge (P3,9),\\ step-by-step recipe guidance (P7,11), recommended lighting for study focus (P9),\\ providing product information being viewed (P2),\\ suggestions for reflection topics after reading (P4),\\ summarizing key concepts of a study page (P8),\\ tracking daily calorie and nutrient intake (P8,10),\\ book location guidance to encourage daily reading (P4),\\ guidance to return books to their original location (P7),\\ alerts for attempting to store frozen items in the refrigerator (P1)\end{tabular} \\ \bottomrule[0.5pt]
\end{tabular}
\end{table*}

\begin{table*}[]
\begin{tabular}{lll}
\toprule[1pt]
\textbf{\begin{tabular}[c]{@{}l@{}}Three\\ Roles\end{tabular}} & \textbf{\begin{tabular}[c]{@{}l@{}}Purpose of\\ Developed Features\end{tabular}} & \textbf{Specific Features Developed by Participants} \\ \midrule[1pt]
\multirow{2}{*}{Advisory} & \begin{tabular}[c]{@{}l@{}}To diagnose\\ daily problems and\\ identify their causes\end{tabular} & \begin{tabular}[c]{@{}l@{}}Analysis of risky or unhygienic cooking actions (P4,12),\\ analysis of push-up posture correctness (P2), diagnosis of unusual pet behavior (P7),\\ analysis of actions that may damage clothing during care (P4,7),\\ analysis of environmental factors affecting study focus (P9),\\ suggestions on reasons for frequent stress during cleaning (P4),\\ analysis of causes of sink dirt buildup (P4),\\ analysis of daily activities affecting sleep quality (P10),\\ analysis of unhealthy sleep postures (P12)\end{tabular} \\ \cline{2-3} 
 & \begin{tabular}[c]{@{}l@{}}To provide\\ actionable solutions\\ to solve problems\end{tabular} & \begin{tabular}[c]{@{}l@{}}Guidance on effective organization and tidying (P10),\\ recommendations for better plant care (P5),\\ guidance on effective bathroom stain removal methods (P5),\\ recommendations for nutritional supplements to address dietary deficiencies (P1,6), \\ fashion suggestions based on hairstyle, body type, weather, and trends (P7,11,12),\\ recipe suggestions using available fridge ingredients (P5,10,11),\\ recommendations for optimal washing modes for mixed laundry loads (P3,9,11,12)\end{tabular} \\ \bottomrule[1pt]
\end{tabular}
\end{table*}

\end{document}